\documentclass[12pt,english]{article}
\usepackage{mathtools,float}
\allowdisplaybreaks[4]
\usepackage[dvipsnames]{xcolor}
\usepackage[T1]{fontenc}
\usepackage[latin9]{inputenc}
\usepackage{geometry}
\usepackage{amsmath,amsthm,amssymb}
\usepackage[authoryear]{natbib}
\usepackage{booktabs}
\usepackage{graphicx}
\usepackage{mdframed}
\usepackage[dvipsnames]{xcolor}

 \theoremstyle{definition}
  
   \theoremstyle{plain}
  \newtheorem{assumption}{\protect\assumptionname}
\theoremstyle{plain}
\newtheorem{thm}{\protect\theoremname}
\theoremstyle{plain}

  \theoremstyle{plain}
  
    \theoremstyle{example}
  
    \theoremstyle{plain}
  \newtheorem{lem}{\protect\lemmaname}
    \newtheorem{algo}{\protect\algorithmname}

\theoremstyle{definition}

\definecolor{darkblue}{rgb}{0, 0, 0.5}

\usepackage{hyperref}
\hypersetup{
     colorlinks = true,
     citecolor = Maroon,
     urlcolor = darkblue,
     linkcolor = darkblue
}

\makeatother

  \providecommand{\assumptionname}{Assumption}
  \providecommand{\examplename}{Example}
  \providecommand{\propositionname}{Proposition}
\providecommand{\theoremname}{Theorem}
\providecommand{\remarkname}{Remark}
\providecommand{\corollaryname}{Corollary}
  \providecommand{\lemmaname}{Lemma}
    \providecommand{\algorithmname}{Algorithm}

\global\long\def\RR{\mathbb{R}}

\global\long\def\RR{\mathbb{R}}

\global\long\def\Acal{\mathcal{A}}%

\global\long\def\Ccal{\mathcal{C}}%

\global\long\def\hF{\hat{F}}%

\global\long\def\hG{\hat{G}}%

\global\long\def\tG{\tilde{G}}%

\global\long\def\hL{\hat{L}}%

\global\long\def\Mcal{\mathcal{M}}%

\global\long\def\RR{\mathbb{R}}%

\global\long\def\Tcal{\mathcal{T}}%

\global\long\def\hU{\hat{U}}%

\global\long\def\hV{\hat{V}}%

\global\long\def\oneb{\mathbf{1}}%

\global\long\def\Xcal{\mathcal{X}}%

\global\long\def\hQ{\hat{Q}}%

\global\long\def\hb{\hat{b}}%

\global\long\def\tV{\tilde{V}}%

\newtheorem{innercustomthm}{Theorem}
\newenvironment{customthm}[1]
  {\renewcommand\theinnercustomthm{#1}\innercustomthm}
  {\endinnercustomthm}
  
  \newtheorem{innercustomass}{Assumption}
\newenvironment{customass}[1]
  {\renewcommand\theinnercustomass{#1}\innercustomass}
  {\endinnercustomass}

  \newtheorem{innercustomlem}{Lemma}
\newenvironment{customlem}[1]
  {\renewcommand\theinnercustomlem{#1}\innercustomlem}
  {\endinnercustomlem}
  
    \newtheorem{innercustomalgo}{Algorithm}
\newenvironment{customalgo}[1]
  {\renewcommand\theinnercustomalgo{#1}\innercustomalgo}
  {\endinnercustomalgo}

\begin{document}

\title{Distributional conformal prediction\footnote{We are grateful to the Editor, two anonymous referees, Dimitris Politis, and Allan Timmermann for valuable comments. W\"uthrich is also affiliated with CESifo and the Ifo Institute. Chernozhukov gratefully acknowledges funding by the National Science Foundation. The usual disclaimer applies.}}

  \author{Victor Chernozhukov\thanks{Massachusetts Institute of Technology; 50 Memorial Drive, E52-361B, Cambridge, MA 02142, USA; Email: \url{vchern@mit.edu}} \qquad Kaspar W\"uthrich\thanks{Department of Economics, University of California San Diego, 9500 Gilman Dr., La Jolla, CA 92093, USA; Email: \url{kwuthrich@ucsd.edu}} \qquad Yinchu Zhu\thanks{Brandeis University; 415 South Street, Waltham, MA 02453, USA; Email: \url{yinchuzhu@brandeis.edu}}}

\date{First version on arXiv: September, 17 2019 \quad This version: \today}

\maketitle

\linespread{1.15} \parskip 0in

\begin{abstract}

We propose a robust method for constructing conditionally valid prediction intervals based on models for conditional distributions such as quantile and distribution regression. Our approach can be applied to  important prediction problems including cross-sectional prediction, $k$-step-ahead forecasts, synthetic controls and counterfactual prediction, and individual treatment effects prediction. Our method exploits the probability integral transform and relies on permuting estimated ranks. Unlike regression residuals, ranks are independent of the predictors, allowing us to construct conditionally valid prediction intervals under heteroskedasticity. We establish approximate conditional validity under consistent estimation and provide  approximate unconditional validity under model misspecification, overfitting, and with time series data. We also propose a simple ``shape'' adjustment of our baseline method that yields optimal prediction intervals.

\medskip


\end{abstract}

\maketitle

\newpage

\linespread{1.25} \parskip .05in

\section{Introduction}

We develop a robust approach for constructing prediction intervals based on models for conditional distributions. The proposed method is generic and can be implemented using a great variety of flexible and powerful methods, including conventional quantile regression (QR) \citep{koenker1978}, distribution regression (DR) \citep[e.g.,][]{foresi1995conditional,chernozhukov2013inference}, as well as non-parametric and high-dimensional machine learning methods such as quantile neural networks \citep[e.g.,][]{taylor2000quantile} and quantile trees and random forests \citep[e.g.,][]{chaudhuri2002,meinshausen2006quantile}.

We observe data $\left\{(Y_t,X_t)\right\}_{t=1}^T$, where $Y_t$ is a continuous outcome of interest and $X_t$ is a $p\times 1$ vector of predictors. Our task is to predict $Y_{T+1}$ given knowledge of $X_{T+1}$. This setting encompasses many classical cross-sectional and time series prediction problems. Moreover, our approach can be applied to synthetic control settings where the goal is to predict counterfactuals in the absence of a policy intervention \citep[e.g.,][]{cattaneo2019prediction,chernozhukov2021conformal} and to the problem of predicting individual treatment effects \citep[e.g.,][]{kivaranovic2020conformal,lei2020conformal}.

With iid (or exchangeable data), standard conformal prediction methods, which are based on modeling the conditional mean, yield prediction intervals $\widehat{\mathcal{C}_{(1-\alpha)}}$ that satisfy
\begin{eqnarray}
P\left(Y_{T+1}\in \widehat{\mathcal{C}_{(1-\alpha)}} \left(X_{T+1}\right)\right)\ge 1-\alpha \label{eq:uncond_coverage}
\end{eqnarray}
for a given miscoverage level $\alpha\in (0,1)$. A prediction interval satisfying this property is said to be \emph{unconditionally} valid. Unconditionally valid prediction intervals guarantee accurate coverage on average, treating $(Y_{T+1},X_{T+1})$ and $\left\{(Y_t,X_t)\right\}_{t=1}^T$ as random. 

However, in many applications, unconditional validity may be unsatisfactory. Let us consider three examples; see \cite{romano2019malice,foygel2021limits} for further examples and discussions. First, from a fairness perspective, data-driven recommendation systems should guarantee equalized coverage across protected groups, in which case the goal is to construct prediction intervals that are valid conditional on a protected attribute such as race or gender \citep[][]{romano2019malice}. Second, as in Section \ref{sec:predicting_stock_returns}, consider the problem of predicting stock returns given the realized volatility. Since the distribution of returns is more dispersed when the variance is higher, a natural prediction algorithm should yield wider prediction intervals for higher values of volatility. That is, the prediction interval should be valid conditional on the known value of realized volatility rather than on average. Third, as in Section \ref{sec:predicting_wages}, suppose our goal is to predict wages based on an individual's education and experience. An unconditionally valid prediction interval exhibits coverage 90\% on average across all individuals but may contain the true wage of high-school dropouts with no work experience with probability zero. A more useful prediction interval should exhibit correct coverage conditional on an individual's observed education and experience and contain the true wage with 90\% probability for every single individual.

Motivated by this discussion, we develop a \emph{distributional conformal prediction} (DCP) method for constructing prediction intervals that are approximately valid conditional on the full vector of predictors $X_{T+1}$, while treating $Y_{T+1}$ and $\left\{(Y_t,X_t)\right\}_{t=1}^T$ as random:
\begin{eqnarray}
P\left(Y_{T+1}\in \widehat{\mathcal{C}_{(1-\alpha)}} \left(X_{T+1}\right)\mid X_{T+1}\right) \ge 1-\alpha + o_P(1). \label{eq:cond_coverage}
\end{eqnarray}
A prediction interval satisfying property \eqref{eq:cond_coverage} as $T\rightarrow \infty$ is said to be approximately \emph{conditionally} valid.\footnote{See, for example, \cite{lei2014distribution,sesia2019comparison,foygel2021limits} for a further discussion of the difference between conditional and unconditional validity.}

While the requirement in \eqref{eq:cond_coverage} is natural in many applications, there are also other notions of conditional validity. Instead of conditioning on $X_{T+1}$ (object conditional), one can also study the conditional coverage probability given the training sample $\left\{(Y_t,X_t)\right\}_{t=1}^T$ (training conditional) or given $Y_{T+1}$ (label conditional) or combinations of them; see \cite{vovk2012conditional} for a detailed discussion. By Proposition 2 of \cite{vovk2012conditional},  inductive conformal predictions (also known as split-sample conformal predictions) automatically achieve training conditional validity as long as the training sample is large enough. In classification problems (the support of $Y_{T+1}$ is a finite set), label conditional validity is often of great interest as it is important to know the error rates for different categories and provides useful information on false positive and false negative rates \citep{vovk2012conditional}.  In  \cite{vovk2012conditional}, label conditional validity is achieved by forming the conformity score within each category. Both training and label conditional validity can be achieved in a distribution-free way, i.e., for a given procedure, the conditional validity holds for any distribution of the data.

However, object conditional validity in the sense of   \eqref{eq:cond_coverage} cannot be achieved in a distribution-free way for non-trivial predictions.  By \cite{{vovk2012conditional,lei2014distribution,foygel2021limits}}, any prediction set satisfying  \eqref{eq:cond_coverage} for every probability distribution of $(X_t,Y_t)$ has infinite Lebesgue measure with non-trivial probability. Therefore, we only aim to achieve \eqref{eq:cond_coverage}  for a limited class of probability distributions. The construction of the proposed prediction set $ \widehat{\mathcal{C}_{(1-\alpha)}}$ relies on learning the conditional distribution $Y_t\mid X_t$ and  we only hope for conditional validity in \eqref{eq:cond_coverage} in the class of distributions that can be learned well. In particular, this  class of distributions are those satisfying our regularity conditions.

Our empirical results demonstrate the importance of using DCP instead of standard conformal prediction methods based on modeling the conditional mean. When predicting daily stock returns in Section \ref{sec:predicting_stock_returns}, the coverage probability of the 90\% mean-based conformal prediction interval can drop to around 50\% when the realized volatility is high. By contrast, DCP provides a coverage probability close to 90\% for all values of realized volatility. This finding is important since volatility tends to be high during periods of crisis when accurate risk assessments are most needed. When predicting wages in Section \ref{sec:predicting_wages}, we find that the DCP prediction intervals contain the true wage with probability close to 90\% for most individuals, whereas standard mean-based conformal prediction intervals either substantially under- or overcover.

To motivate DCP, note that a conditionally valid prediction interval is given by
\begin{equation}
\left[Q\left(\frac{\alpha}{2} , x\right),Q\left(1-\frac{\alpha}{2}, x \right)\right], \label{eq:ideal_prediction_interval}
\end{equation}
where $Q(\tau, x)$ is the $\tau$-quantile of $Y_t$ given $X_t=x$. To implement the prediction interval \eqref{eq:ideal_prediction_interval}, a plug-in approach would replace $Q$ with a consistent estimator $\hat{Q}$
\begin{equation}
\left[\hat{Q}\left(\frac{\alpha}{2},x \right),\hat{Q}\left(1-\frac{\alpha}{2},x \right)\right]. \label{eq:naive_prediction_interval}
\end{equation}
This approach exhibits two well-known drawbacks. First, it will often exhibit undercoverage in finite samples \citep[e.g.,][]{romano2019conformalized}. Second, it is neither conditionally nor unconditionally valid under misspecification. 

We build upon conformal prediction \citep{vovk2005algorithmic,vovk2009online} and use the conditional ranking as a conformity score. This choice is particularly useful when working with regression models for conditional distributions such as QR and DR.\footnote{This transformation is also very useful in other prediction problems \citep[e.g.,][]{politis2015modelfree}.} Our method is conditionally valid under correct specification, while the construction of the procedure as a conformal prediction method guarantees the unconditional validity under misspecification. Let $F(y,x)=P(Y_t\leq y \mid X_t=x)$ denote the conditional cumulative distribution function (CDF) of $Y_t$ given $X_t=x$. Throughout the paper, we assume that $F(\cdot,X_t)$ is a continuous function almost surely. Our method is based on the probability integral transform, which states that the \emph{conditional rank}, $U_t:=F\left( Y_t,X_t\right)$, has the uniform distribution on $(0,1)$ and is independent of $X_t$.

To construct the prediction interval, we test the plausibility of each $y \in \mathbb{R}$. By the probability integral transform, conditional on $X_{T+1}$, $F(Y_{T+1},X_{T+1}) $ belongs to $[\alpha/2,1-\alpha/2] $ with probability $1-\alpha$. Thus, collecting all values $y\in \mathbb{R}$ satisfying $F(y,X_{T+1})\in[\alpha/2,1-\alpha/2] $ yields a conditionally valid prediction interval in the sense of \eqref{eq:cond_coverage}. We operationalize this idea by proposing a conformal prediction procedure based on the estimated ranks, $\hat{U}_{t}^{(y)}:=\hat{F}^{(y)}(Y_t,X_t)$. For each $y \in \mathbb{R}$, $\hat{F}^{(y)}$ is an estimator of $F$ obtained based on the augmented data, $\{(Y_t,X_t)\}_{t=1}^{T+1}$, where $Y_{T+1}=y$. Data augmentation is a key feature of conformal prediction. It implies the model-free unconditional exact finite-sample validity with iid (or exchangeable) data and, thus, guards against model misspecification and overfitting. Without data augmentation, the resulting prediction intervals are not exactly valid, not even with correct specification and iid data.

Our baseline method asymptotically coincides with the oracle interval in \eqref{eq:ideal_prediction_interval}. 
This  oracle interval may not be the shortest possible prediction interval in general. Therefore, we also develop a simple and easy-to-implement adjustment of our baseline method for improving efficiency, which we refer to as \emph{optimal DCP}.  In Section \ref{sec:predicting_wages}, we show empirically that optimal DCP yields shorter prediction intervals than baseline DCP when the conditional distribution is skewed.

We establish the following theoretical performance guarantees for the baseline and optimal DCP. 
\begin{enumerate}\setlength{\itemsep}{0pt}\setlength{\parskip}{0pt}
\item[(i)] Asymptotic conditional validity under consistent estimation of the conditional CDF
\item[(ii)] Unconditional validity under model misspecification:
\vspace{-5pt} 
\begin{itemize}\setlength{\itemsep}{0pt}\setlength{\parskip}{0pt}
\item[(a)] Finite-sample validity with iid\ (or exchangeable) data 
\item[(b)] Asymptotic validity with time series data
\end{itemize}
\item[(iii)] For optimal DCP:
\begin{itemize}\setlength{\itemsep}{0pt}\setlength{\parskip}{0pt}
\item[(a)] Under weak conditions: asymptotic conditional validity and optimality (shortest length)
\item[(b)] Under strong conditions: asymptotic convergence to the optimal prediction interval
\end{itemize}
\end{enumerate}
\subsection{Motivating Example}
\label{sec:intro_conformal_prediction}

We illustrate the advantages of DCP relative to mean-based conformal prediction \citep[e.g.,][]{lei2018distributionfree} based on the following simple analytical example. 
\begin{equation}
Y_t=X_t+X_t\varepsilon_t, \quad X_t\overset{iid}\sim \text{Uniform}(0,1), \quad \varepsilon_t\overset{iid}\sim N(0,1).\label{eq:dgp_motivating_example}
\end{equation}
Our motivating example draws on \cite{koenker1982robust,koenker2005book,lei2018distributionfree,romano2019conformalized}. We focus on the population conformal prediction (or oracle) problem under correct specification and abstract from finite sample issues.

Mean-based conformal prediction is based on the residuals $R_t=Y_t-E(Y_t\mid X_t)=Y_t-X_t=X_t\varepsilon_t$. The mean-based prediction interval is
\begin{equation}
\mathcal{C}^{\rm reg}_{(1-\alpha)}(x)=\left[x-Q_{|R|}(1-\alpha),x+Q_{|R|}(1-\alpha)\right], \label{eq:interval_regression}
\end{equation}
where $Q_{|R|}(1-\alpha)$ is the $(1-\alpha)$-quantile of the distribution of $|R_{t}|$. An important property and drawback of $\mathcal{C}^{\rm reg}_{(1-\alpha)}$ is that its length, $2 \cdot Q_{|R|}(1-\alpha)$, is fixed and does not depend on $X_{T+1}=x$  \citep{lei2018distributionfree,romano2019conformalized}. This feature implies that $\mathcal{C}^{\rm reg}_{(1-\alpha)}$ is not adaptive to the heteroskedasticity in the location-scale model \eqref{eq:dgp_motivating_example} and not conditionally valid.

DCP is based on the ranks $U_t=\Phi\left(\varepsilon_t \right)$, where $\Phi(\cdot)$ is the CDF of $N(0,1)$. The DCP prediction interval is
\begin{eqnarray}
\mathcal{C}^{\rm dcp}_{(1-\alpha)}(x)
=\left[x-x\cdot Q_{|\varepsilon|}(1-\alpha),x+x\cdot Q_{|\varepsilon|}(1-\alpha)\right], \label{eq:interval_regression}
\end{eqnarray}
where $Q_{|\varepsilon|}(1-\alpha)=\Phi^{-1}(1-\alpha/2)$ is the $(1-\alpha)$-quantile of $|\varepsilon_t|$. Unlike $\mathcal{C}^{\rm reg}_{(1-\alpha)}$, the length of $\mathcal{C}^{\rm dcp}_{(1-\alpha)}$, $2x\cdot Q_{|\varepsilon|}(1-\alpha)$, depends on $X_{T+1}=x$. Our construction automatically adapts to the heteroskedasticity in model \eqref{eq:dgp_motivating_example} and is conditionally valid.

Figure \ref{fig:intro} provides an illustration. Panel (a) shows that the conditional length of $\mathcal{C}^{\rm reg}_{(0.9)}$ is constant, whereas the length of $\mathcal{C}^{\rm dcp}_{(0.9)}$ varies as a function of $x$. $\mathcal{C}^{\rm dcp}_{(0.9)}$ is shorter than $\mathcal{C}^{\rm reg}_{(0.9)}$ for low values and wider for high values of $x$. Panel (b) shows that $\mathcal{C}^{\rm dcp}_{(0.9)}$ is valid for all $x$, whereas $\mathcal{C}^{\rm reg}_{(0.9)}$ overcovers for low values and undercovers for high values of $x$. Figure \ref{fig:intro} illustrates the advantage of our method. For predictor values where the conditional variance is low, it yields shorter prediction intervals, while ensuring conditional coverage for values where the conditional dispersion is large by suitably enlarging the prediction interval.
\begin{figure}[H]
\begin{center}
\includegraphics[width=0.4\textwidth,trim=0 1.25cm 0 1cm]{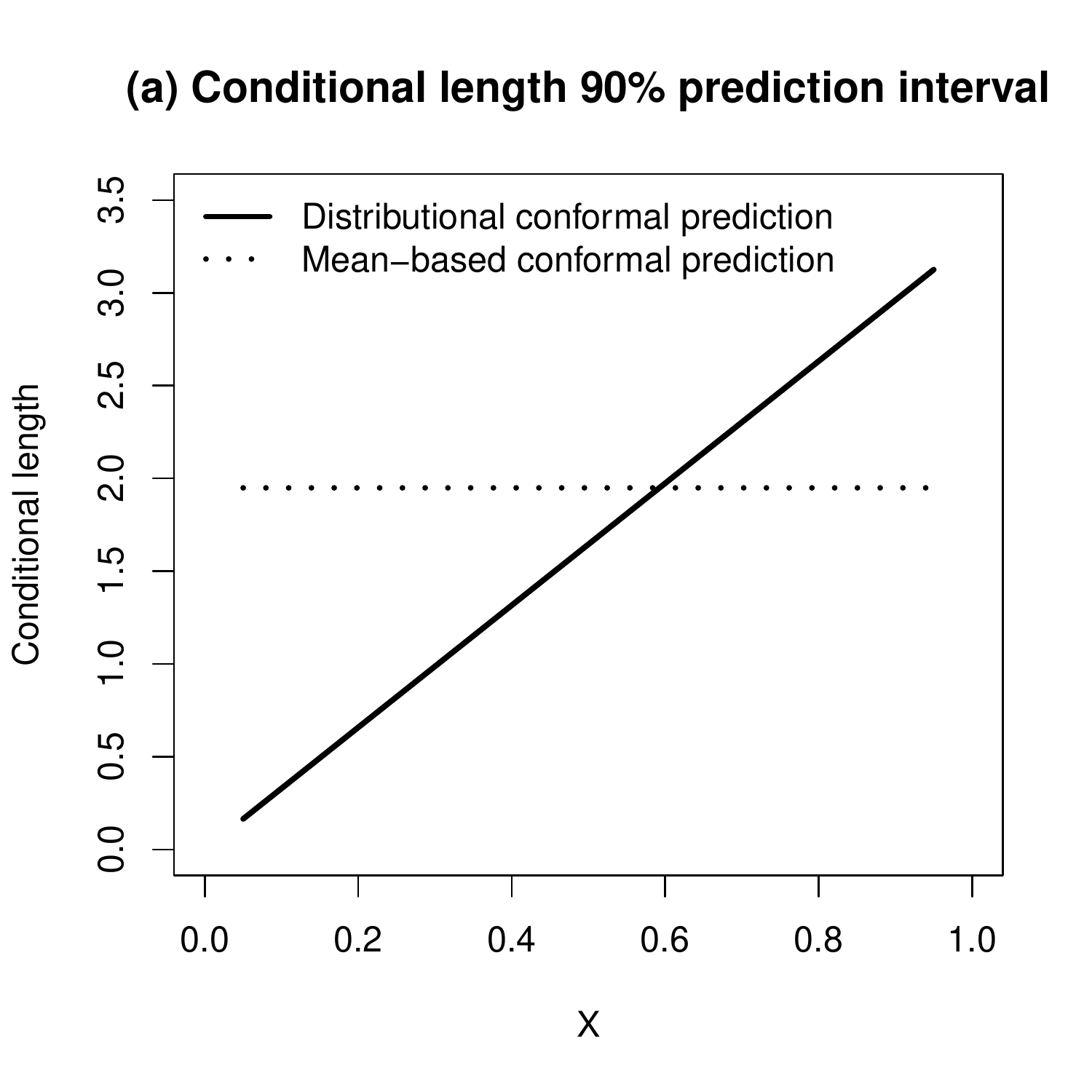}
\includegraphics[width=0.4\textwidth,trim=0 1.25cm 0 1cm]{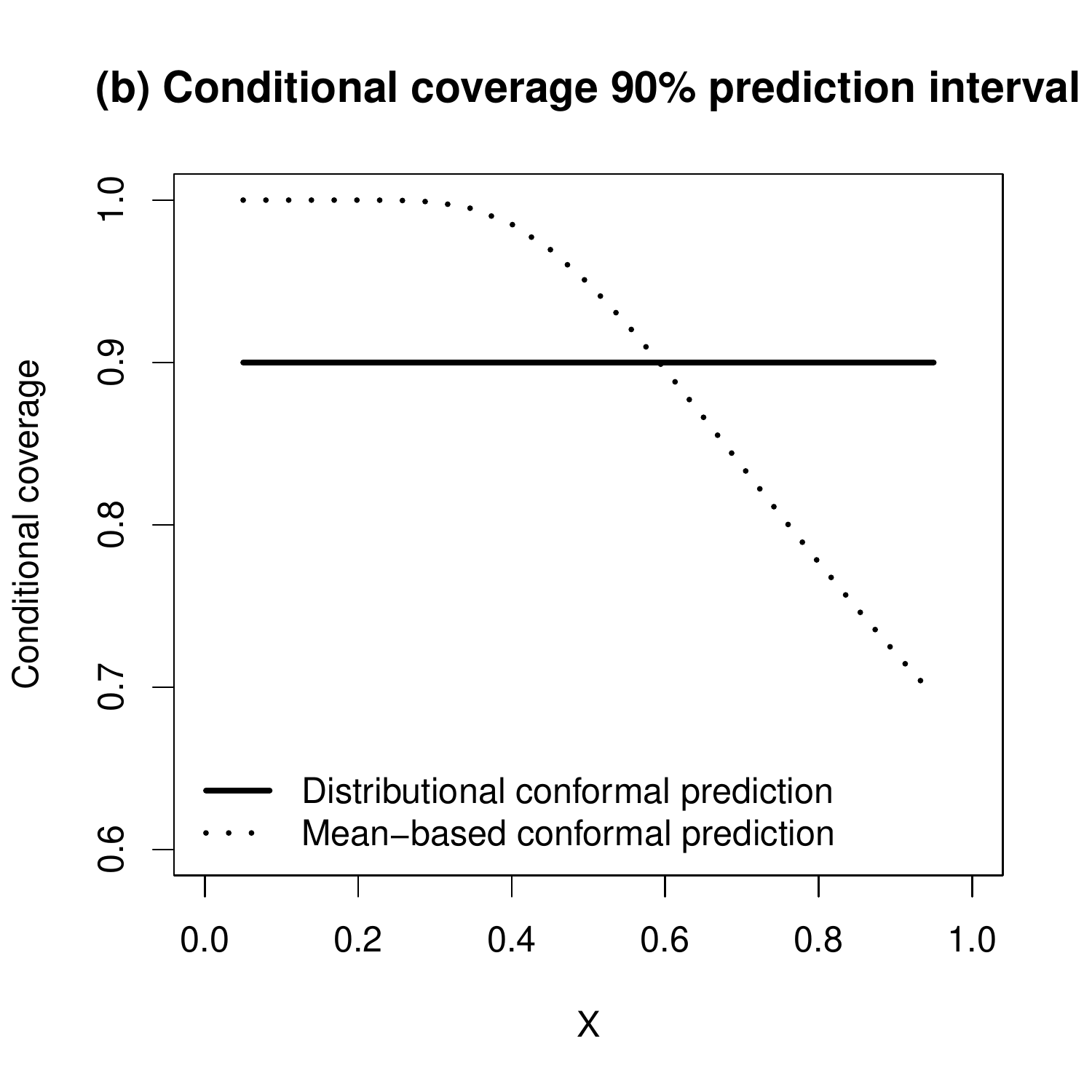}
\caption{Motivating example}
\label{fig:intro}
\end{center}
\end{figure}

\subsection{Related Literature}
\label{sec:literature}
We build on and contribute to the literature on conformal prediction \citep[e.g.,][]{vovk2005algorithmic,vovk2012conditional,vovk2009online,lei2013distribution,lei2014distribution,lei2018distributionfree,chern2018COLT,romano2019conformalized}, the literature on model-free prediction \citep{politis2013modelfree,politis2015modelfree}, as well as the literature on quantile prediction methods \citep[see, e.g.,][for a review]{komunjer2013handbook}.

Within the conformal prediction literature, our paper is most closely related to \cite{lei2014distribution}, \cite{lei2018distributionfree}, and \cite{romano2019conformalized}. \cite{lei2014distribution} propose conditionally valid and asymptotically efficient conformal prediction intervals based on estimators of the conditional density. We take a different and complementary approach, allowing researchers to leverage powerful regression methods for modeling conditional distributions, including QR and DR approaches. \cite{lei2018distributionfree} develop conformal prediction methods based on regression models for conditional expectations. However, as discussed in Section \ref{sec:intro_conformal_prediction}, this approach is not conditionally valid under heteroskedasticity. They also propose a locally weighted conformal prediction approach, where the regression residuals are weighted by the inverse of a measure of their variability. This approach can alleviate some of the limitations of mean-based conformal prediction but is motivated by and based on restrictive locations-scale models. By contrast, our approach is generic and exploits flexible and substantially more general models for the whole conditional distribution. 

\cite{romano2019conformalized} propose a split conformal approach based on QR models, which they call conformalized quantile regression (CQR). See also \cite{sesia2019comparison,kivaranovic2020adaptive} for related approaches and \cite{vovk2020conformal} for a general approach to adaptive conformal prediction. Their approach is based on splitting the data into two subsets, $\mathcal{T}_1$ and $\mathcal{T}_2$. Based on $\mathcal{T}_1$, they estimate two separate quantile functions $\hat{Q}(\alpha/2,x)$ and $\hat{Q}(1-\alpha/2,x)$ and construct the prediction intervals as
\[
\left[ \hat{Q}(\alpha/2,x)-Q_{E}(1-\alpha),\hat{Q}(1-\alpha/2,x)+Q_{E}(1-\alpha)\right],
\]
where $Q_{E}(1-\alpha)$ is the $(1-\alpha)(1+1/|\mathcal{T}_2|)$-th empirical quantile of $$E_t=\max\left\{\hat{Q}(\alpha/2,X_t)-Y_t,Y_t- \hat{Q}(1-\alpha/2,x)\right\}$$ in $\mathcal{T}_2$. Constructing prediction intervals based on deviations from quantile estimates is similar to working with deviations from mean estimates, as the deviations are measured in absolute levels. By contrast, exploiting the probability integral transform, our approach is generic and relies on permuting ranks, which naturally have the same scaling on $(0,1)$. Note, however, that our paper was inspired by \cite{romano2019conformalized} and we view our proposal as a (fully quantile-rank based) refinement of \cite{romano2019conformalized}.

Our adjustment for constructing efficient prediction intervals is related to and inspired by conformal prediction literature on minimum-volume prediction sets based on density estimators \citep[e.g.,][]{lei2013distribution,lei2014distribution,eck2019efficient,izbicki2019flexible,izbicki2020cd}  and nearest-neighbor estimators \cite{gyorfi2020nearest}. It is most closely related and can be viewed as an alternative to conformal histogram regression \citep{sesia2021conformal}. The main differences between our approach and conformal histogram regression are the following. First, our method is based on an optimization problem formulated in terms of estimated quantile functions and does not require estimating a conditional density or histogram. Second, we do not work with nested sets but instead use a simple adjustment of our baseline conformity score. Finally, our approach works for general outcome distributions and does not rely on assuming unimodal distributions.

Conceptually, our paper is further related to the transformation-based model-free prediction approach developed in \cite{politis2013modelfree} and \cite{politis2015modelfree} in that we rely on transformations of the original setup into one that is easier to work with (i.e., ranks which are uniformly distributed) and study the properties of our approach in a model-free setting. An important difference is the implementation of the resulting procedure. The transformation-based approach is based on the bootstrap, whereas our approach is based on permuting ranks. Permuting ranks estimated based on the augmented data guarantees the model-free finite sample validity of our method with exchangeable data. To our knowledge, no exact finite-sample validity results have been developed for the bootstrap-based approach.

\section{Distributional Conformal Prediction}
Here we introduce DCP. We present a full and a split sample version of our method.

\subsection{Full Distributional Conformal Prediction}
Let $y$ denote a test value for $Y_{T+1}$. We test plausibility of each value $y \in \mathbb{R}$, collect all plausible values, and report them as the prediction set. In practice, we consider a grid of test values $\mathcal{Y}_{\rm trial}$.\footnote{For example, we can choose $\mathcal{Y}_{\rm trial}$ to be a fine grid between $-\max_{1\le t \le T} |Y_t|$ and $\max_{1\le t \le T} |Y_t|$. This choice has a theoretical justification since, under exchangeability, $P\left(|Y_{T+1}|\ge \max_{1\le t \le T} |Y_t|\right)\le 1/(1+T)$ \citep{chen2016trimmed}; see also the discussion in the \texttt{conformalInference} \texttt{R}-package (\url{https://github.com/ryantibs/conformal}).} Define the augmented data $Z^{(y)}=\{Z_t^{(y)}\}_{t=1}^{T+1}$, where 
\begin{equation}\label{eq: def Z}
Z_t^{(y)}=\begin{cases}
(Y_t,X_t) & \textrm{if}\ 1\leq t\leq T \\
(y,X_t) & \textrm{if}\ t=T+1
\end{cases}
\end{equation}

Based on the augmented dataset $Z^{(y)}$, we estimate the conditional CDF using a suitable method such as QR and DR, which are discussed in more detail in the SI Appendix. Let $\hat{F}^{(y)}$ denote the estimator for $F$ based on the augmented sample. If the original estimate is not monotonic, we rearrange it \citep[e.g.,][]{chernozhukov2009improving,chernozhukov2010quantile} so that $\hat{F}^{(y)}(\cdot,x)$ is always monotonic. To simplify the exposition, we keep these rearrangements implicit.

We compute the ranks $\{\hat{U}^{(y)}_t\}_{t=1}^{T+1}$, where
\begin{equation}\label{eq:augmented_data}
\hat{U}^{(y)}_t=\begin{cases}
\hat{F}^{(y)}(Y_t,X_t)& \textrm{if}\ 1\leq t\leq T \\
\hat{F}^{(y)}(y, X_t)& \textrm{if}\ t=T+1
\end{cases}
\end{equation}
and obtain $p$-values as
\begin{equation}
\hat{p}(y)=\frac{1}{T+1}\sum_{t=1}^{T+1}\oneb\left\{\hV^{(y)}_{t}\ge \hV^{(y)}_{T+1} \right\}, \label{eq:p_value}
\end{equation}
where $\hV^{(y)}_{t}:=\psi(\hU^{(y)}_{t})$, and $\psi(\cdot)$ is a deterministic function. For our baseline method, we use $\psi(x)=|x-1/2|$. 
In Section \ref{sec:extension}, we show how to choose $\psi$ optimally to ensure efficiency.  Prediction intervals are computed as $\widehat{\mathcal{C}^{\rm full}_{(1-\alpha)}}(X_{T+1})=\left\{y\in \mathcal{Y}_{\rm trial}:\hat{p}(y)>\alpha \right\}$.\footnote{Instead of $\widehat{\mathcal{C}^{\rm full}_{(1-\alpha)}}(X_{T+1})$ we typically report the closed interval
    $\widetilde{\mathcal{C}^{\rm full}_{(1-\alpha)}}(X_{T+1})=\left[\min\left(\widehat{\mathcal{C}^{\rm full}_{(1-\alpha)}}(X_{T+1})\right),\max \left( \widehat{\mathcal{C}^{\rm full}_{(1-\alpha)}}(X_{T+1})\right)\right]$.}
 We summarize our approach in Algorithm \ref{algo:conformal}.

\begin{algo}[Full DCP]
\label{algo:conformal}
\text{ }
\vspace{-2mm}
\begin{itemize} \setlength{\itemsep}{0pt}\setlength{\parskip}{0pt}
\item[] \textbf{Input:} Data $\left\{(Y_t,X_t)\right\}_{t=1}^T$, miscoverage level $\alpha \in (0,1)$, a point $X_{T+1}$, test values $\mathcal{Y}_{\rm trial}$
\item[] \textbf{Process:}
For $y\in \mathcal{Y}_{\rm trial}$,
\begin{enumerate} \setlength{\itemsep}{0pt}\setlength{\parskip}{0pt}
\item define the augmented data $Z^{(y)}$ as in \eqref{eq:augmented_data}
\item compute $\hat{p}(y)$ as in \eqref{eq:p_value}
\end{enumerate}
\item[] \textbf{Output:} Return $(1-\alpha)$ prediction set $\widehat{\mathcal{C}^{\rm full}_{(1-\alpha)}}(X_{T+1})=\left\{y\in \mathcal{Y}_{\rm trial}:\hat{p}(y)>\alpha \right\}$
\end{itemize}
\end{algo}

\subsection{Split Distributional Conformal Prediction}

An important drawback of full DCP (Algorithm \ref{algo:conformal}) is its computational burden due to the grid search. Since $\hat{F}^{(y)}$ is obtained based on the augmented data, one has to choose $\mathcal{Y}_{\rm trial}$ and re-estimate the entire conditional distribution for all $y\in \mathcal{Y}_{\rm trial}$. Therefore, we propose a split conformal procedure that exploits sample splitting, avoids grid search, and only requires estimating $F$ once. Sample splitting is a popular approach for improving the computational performance of conformal prediction methods \citep[e.g.,][]{lei2018distributionfree,romano2019conformalized}. 

\begin{algo}[Split DCP]
\label{algo: split dcp}
\text{ }
\begin{itemize} \setlength{\itemsep}{0pt}\setlength{\parskip}{0pt}
\item[] \textbf{Input:} Data $\left\{(Y_t,X_t)\right\}_{t=1}^T$, miscoverage level $\alpha \in (0,1)$, point $X_{T+1}$
\item[] \textbf{Process:}
\begin{enumerate} \setlength{\itemsep}{0pt}\setlength{\parskip}{0pt}
\item Split $\{1,\dots,T\}$ into $\mathcal{T}_1:=\{1,\dots,T_0\}$ and $\mathcal{T}_2:=\{T_0+1,\dots,T\}$
\item Obtain $\hat{F}$ based on $\{Z_t\}_{t\in \mathcal{T}_1}$
\item Compute $\{\hat{V}_t\}_{t\in \mathcal{T}_2}=\{\psi(\hat{U}_t)\}_{t\in \mathcal{T}_2}$, where $\hat{U}_t=\hat{F}(Y_t,X_t)$.
\item Compute $\hat{Q}_{\mathcal{T}_2}$, the $(1-\alpha)(1+1/|\mathcal{T}_2|)$ empirical quantile of $\{\hat{V}_t\}_{t\in \mathcal{T}_2}$.
\end{enumerate}
\item[] \textbf{Output:} Return $(1-\alpha)$ prediction set $\widehat{\mathcal{C}^{\rm split}_{(1-\alpha)}}(X_{T+1})=\left\{y:\psi\left( \hat{F}(y,X_{T+1})\right)\le \hat{Q}_{\mathcal{T}_2}\right\}$. 

(Since $\hat{F}(\cdot, X_{T+1})$ is monotonic, $\widehat{\mathcal{C}^{\rm split}_{(1-\alpha)}}(X_{T+1})$ is an interval.)
\end{itemize}
\end{algo}

In Algorithm \ref{algo: split dcp}, we split $\{1,\dots,T\}$ into $\{1,\dots,T_0\}$ and $\{T_0+1,\dots,T\}$. With iid data, one can also consider random splits.

Split DCP lends itself naturally to simple in-sample validity checks with both cross-sectional and time series data as illustrated in Section \ref{sec:applications}.

\section{Theoretical Performance Guarantees}
\label{sec:theory}

In this section, we establish the theoretical properties of our procedure. We focus on full-sample DCP (Algorithm \ref{algo:conformal}). For the split-sample approach (Algorithm \ref{algo: split dcp}), we provide a modified version (Algorithm \ref{algo: optimal}) in the SI Appendix and present its theoretical properties in Section \ref{sec:extension}.

When the data are iid\ (or exchangeable), our method achieves finite-sample unconditional validity in a model-free manner, as a consequence of general results on conformal inference and permutation inference more generally \citep[e.g.,][]{vovk2005algorithmic,hoeffding1952large}. 
\begin{thm}[Finite sample unconditional validity]\label{thm:finite_sample} Suppose that the data are iid or exchangeable and that the estimator  of the conditional distribution is invariant to permutations of the data. Then
\[
P\left(Y_{T+1}\in \widehat{\mathcal{C}_{(1-\alpha)}^{\rm full}}\left(X_{T+1} \right)\right)\geq 1-\alpha.
\]
\end{thm}
The proof of Theorem \ref{thm:finite_sample} is standard and omitted. Theorem \ref{thm:finite_sample} highlights the strengths and drawbacks of conformal prediction methods. Most commonly-used estimators of the conditional CDF such as QR and DR are invariant to permutations of the data. As a result, Theorem \ref{thm:finite_sample} provides a model-free unconditional performance guarantee in finite samples, allowing for arbitrary misspecification of the model of the conditional CDF. On the other hand, it has a major theoretical drawback. Even with iid data, it provides no guarantee at all on conditional validity. 

Our next theoretical results provide a remedy.  We impose the following weak regularity conditions.
 
\begin{assumption}
	\label{assu: basic} Suppose that there exists a non-random function $F^*(\cdot,\cdot)$ such that the following conditions hold as $T\rightarrow \infty$. Define $V_t:=\psi(F^*(Y_t,X_t))$ for $1\leq t\leq T+1$. 
\begin{enumerate}\setlength{\itemsep}{0pt}\setlength{\parskip}{0pt}
\item There exists a strictly increasing continuous function $\phi:[0,\infty)\rightarrow[0,\infty)$
	such that $\phi(0)=0$ and $(T+1)^{-1}\sum_{t=1}^{T+1}\phi(|\hV_{t}-V_{t}|)=o_{P}(1)$
	and $\hV_{T+1}=V_{T+1}+o_{P}(1)$, where $\hV_{t}:=\hV_{t}^{(Y_{T+1})}=\psi(\hF^{(Y_{T+1})}(Y_t,X_t))$ for  $1\leq t\leq T+1$.
\item $\sup_{v\in\RR}|\tG(v)-G(v)|=o_{P}(1)$, where $\tG(v)=(T+1)^{-1}\sum_{t=1}^{T+1}\oneb\{V_{t}< v\}$ and $G(\cdot)$ is the distribution function of $V_{T+1}$.
\item $\sup_{x_{1}\neq x_{2}}|G(x_{1})-G(x_{2})|/|x_{1}-x_{2}|$ is bounded. 
\end{enumerate}	
\end{assumption}

Assumption \ref{assu: basic} allows for some flexibility with respect to the model estimator. Here, we only require $F^*$ to be a non-random function, which may or may not be $F$. The interpretation is straight-forward when $F^*=F$ since this simply means that the estimator $\hF$ is consistent for $F$. We discuss the case of $F^*\neq F$ after Theorem \ref{thm: unconditional validity asy} below. Note that we can replace the consistency requirement in Assumption  \ref{assu: basic} with a stronger uniform consistency requirement, $\sup_{x,y}|\hF(y,x)-F^\ast(y,x)|=o_P(1)$. 

We also notice that the quantities $\hV_{t}$ and $V_{t}$ are defined under the true $Y_{T+1}$. This means that $\hF^{(y)}$ uses $y=Y_{T+1}$. In other words, the estimator $\hF$ based on the sample $\{(X_t,Y_t) \}_{t=1}^{T+1}$ would be consistent for some $F^*$ if $Y_{T+1}$ were observed.\footnote{This is not really much different from assuming that  $\hF$ based on the sample $\{(X_t,Y_t) \}_{t=1}^{T}$ is consistent for some $F^*$.} Since the goal of Assumption  \ref{assu: basic}  is to guarantee the coverage probability for $Y_{T+1}$, the conditions in Assumption \ref{assu: basic} only need to hold for $y=Y_{T+1}$.

Notice that $\hat{F}$ is consistent for $F^*$ under a very weak norm, and no rate condition is required. 
When $\psi(x)=|x-1/2|$, a simple example of $\phi(\cdot) $ in Assumption \ref{assu: basic}  is $\phi(x)=x^q $ for some $q>0$; in other words, a sufficient condition is $(T+1)^{-1}\sum_{t=1}^{T+1}|\hat{F}(Y_t,X_t)-F^\ast(Y_t,X_t)|^{q}=o_{P}(1)$, which can be verified for many existing estimators with $q=2$.

The following lemma gives the basic consistency result.
\begin{lem}
	\label{lem: basic}Let Assumption \ref{assu: basic} hold. Then $\hG(\hV_{T+1})=G(V_{T+1})+o_{P}(1)$, where  $\hG(v)=(T+1)^{-1}\sum_{t=1}^{T+1}\oneb\{\hV_{t}< v\}$. 
\end{lem}

By Assumption  \ref{assu: basic}, $G(\cdot)$ is uniformly continuous and thus continuous. Since $G(\cdot)$ is the distribution function of $V_{T+1}$, we have that $G(V_{T+1})$ has the uniform distribution on $(0,1)$, i.e., $P(G(V_{T+1})\leq\alpha)=\alpha$. This implies the unconditional asymptotic validity.
\begin{thm}[Asymptotic unconditional validity]
	\label{thm: unconditional validity asy}Let Assumption \ref{assu: basic}
	hold. Then 
	\[
	P\left(Y_{T+1}\in\widehat{\mathcal{C}^{\rm full}_{(1-\alpha)}}\left(X_{T+1} \right)\right)=1-\alpha+o(1).
	\]
\end{thm}

Theorem \ref{thm: unconditional validity asy} establishes the asymptotic unconditional validity of the procedure. Since Theorem \ref{thm:finite_sample} already establishes the unconditional validity in finite-samples for iid or exchangeable data without assuming any consistency of $\hF$, the main purpose of Theorem \ref{thm: unconditional validity asy} is to address the case of non-exchangeable data (e.g., time series data with ergodicity), especially when the model is misspecified (i.e., if $F^*\neq F$).

To illustrate model misspecification, consider the popular linear QR model, which assumes $Q(\tau,x)=x^{\top}\beta(\tau)$ and thus $F(y,x)=F(y,x;\beta)=\int_{0}^{1} \oneb\{x^{\top}\beta(\tau) \leq y \} d\tau$. This model is typically estimated by $\hat{\beta}(\tau)=\arg\min_{\beta}\sum_{t=1}^{T+1}\rho_{\tau}(Y_t-X_t^{\top} \beta)$ with $\rho_{\tau}(a)=a(\tau-\oneb\{a< 0\})$. Under misspecification ($Q(\tau,x)\neq x^{\top}\beta(\tau)$), $\hat{\beta}(\tau)$ is still estimating $\beta^*(\tau)=\arg\min_{\beta}\sum_{t=1}^{T+1}E\rho_{\tau}(Y_t-X_t^{\top} \beta)$ and  $F^*$ is defined using $\beta^*(\cdot)$, e.g., $F^*(y,x)=\int_{0}^{1} \oneb\{x^{\top}\beta^*(\tau) \leq y \} d\tau$. For parametric models, $F^*$ is usually the probability limit of $\hF$. In general, we can consider a model $\mathcal{F}$ and minimize the empirical risk  $\hF=\arg\min_{g\in\mathcal{F}} \sum_{t=1}^{T+1} L(Y_t,X_t,g) $ for some loss function $L$. Even if the model is misspecified ($F\notin \mathcal{F}$), it is still possible to show that $\hF$ is close (in some norm) to $F^*=\arg\min_{g\in\mathcal{F}} \sum_{t=1}^{T+1}E[ L(Y_t,X_t,g) ]$. In the SI Appendix, we provide a more detailed discussion of this and some theoretical results verifying the consistency requirement in  Assumption  \ref{assu: basic} for the time series case; see also \cite{chern2018COLT} for a general discussion of conformal prediction in time series settings.  

The cost of allowing for misspecification is that one cannot guarantee conditional validity when $F^*\neq F$. On the other hand,  Lemma \ref{lem: basic} implies that the prediction intervals are conditionally valid when $F^*=F$.

\begin{thm}[Asymptotic conditional validity] 
	\label{thm: conditional validity asy}Let Assumption \ref{assu: basic}
	hold with $F^*=F$.
	Then
	\[
	P\left(Y_{T+1}\in\widehat{\mathcal{C}^{\rm full}_{(1-\alpha)}}\left( X_{T+1}\right)\mid X_{T+1}\right)=1-\alpha+o_{P}(1).
	\]
\end{thm}

Theorems \ref{thm: unconditional validity asy}--\ref{thm: conditional validity asy} establish the asymptotic validity of our procedure under weak and easy-to-verify conditions. They formalize the key intuition that conditional validity hinges on the quality of the estimator $\hat{F}$ of the conditional CDF.\footnote{In Theorem \ref{thm: conditional validity asy}, we assume $F^*=F$. Since the first version of this paper was written, \cite{candes2021conformalized} have provided more general results where $F^*\approx F$.}

\section{Extension: Optimal DCP}
\label{sec:extension}

In Section \ref{sec:theory}, we have seen that a generic conformity score $\psi(y,x)=|F(y,x)-1/2|$  leads to conditional validity if the conditional distribution $F$ can be estimated consistently. We now characterize an optimal choice of conformity score that results in the shortest prediction interval. Detailed implementation algorithms, technical assumptions, and proofs are provided in the SI Appendix.

Let $\mathcal{Z}$ and $\Xcal$ denote the support of $Z_t=(Y_t,X_t)$ and $X_t$, respectively. The optimal prediction interval is
\begin{equation}
\Ccal^{{\rm opt}}_{(1-\alpha)}(x)=[r_{1}(x,\alpha),\ r_{2}(x,\alpha)],\label{eq: opt con int Y}
\end{equation}
where the functions $r_{1}(\cdot,\cdot),r_{2}(\cdot,\cdot)$ satisfy that for any $x\in\Xcal$, 
\begin{equation}\label{eq: key opt}
r_{2}(x,\alpha)-r_{1}(x,\alpha)=\underset{ F(z_2,x)-F(z_1,x)\geq1-\alpha}{\min}\ z_2-z_1.
\end{equation}

The question is whether it is possible to design a conformity score that achieves the above optimal prediction interval. To answer this question formally, we consider  a generic conformity score $\psi(y,x)$, which might contain components that need to be estimated.

Permuting a large number of values of $\psi(Y_{t},X_{t})$ in conformal predictions amounts to taking the sample $(1-\alpha)$-quantile of $\psi(Y_{t},X_{t})$; for example, following Algorithm \ref{algo: split dcp}, one would $(1-\alpha)(1+1/|\mathcal{T}_2|)$ empirical quantile of $\psi(Y_{t},X_{t})$.  Assuming a law of large numbers, this empirical quantile would be close to the population $(1-\alpha)$-quantile of $\psi(Y_{t},X_{t})$, leading to  the asymptotic conformal prediction interval for $Y_{T+1}$ 
\begin{equation}
\Ccal^{{\rm conf}}_{(1-\alpha)}(X_{T+1})=\{y:\ \psi(y,X_{T+1})\leq Q_\psi(1-\alpha)\},\label{eq: conf int}
\end{equation}
where $Q_\psi(1-\alpha)$ is the $(1-\alpha)$-quantile of $\psi(Y_{t},X_{t})$. The following result shows how to construct the optimal conformity score $\psi$.

\begin{lem}
\label{lem: opt score}
Let $\psi_*(y,x)=|F(y,x)-b(x,\alpha)-(1-\alpha)/2|$, where $b(\cdot,\cdot)$ is a function satisfying that for any $x\in\Xcal$,
\begin{equation}\label{eq: b fun condition}
b(x,\alpha)\in\underset{z\in[0,\alpha]}{\arg\min}\  Q(z+1-\alpha,x)-Q(z,x).
\end{equation}
Let $\Ccal^{{\rm conf}}_{(1-\alpha)}(X_{T+1})$ be defined as in \eqref{eq: conf int} with the above conformity score $\psi_*$.
Assume that $F(\cdot,x)$ is a continuous function for any $x\in\Xcal$. Then $Q_\psi(1-\alpha)=(1-\alpha)/2$ and
$$
\mu\left(\Ccal^{{\rm opt}}_{(1-\alpha)}(X_{T+1})\right)=\mu\left(\Ccal^{{\rm conf}}_{(1-\alpha)}(X_{T+1})\right)\  {\rm almost\ surely},
$$
where $\mu(\cdot)$ denotes the Lebesgue measure. 
If the optimization problem in \eqref{eq: opt con int Y} has a unique solution for any $x\in\Xcal$, then
$$
\Ccal^{{\rm opt}}_{(1-\alpha)}(X_{T+1})=\Ccal^{{\rm conf}}_{(1-\alpha)}(X_{T+1})\  {\rm almost\ surely}.
$$

\end{lem}

Lemma \ref{lem: opt score} motivates conformity scores of the form $\psi_*(y,x)=\left|F(y,x)-[b(x,\alpha)+(1-\alpha)/2]\right|$, where $b(\cdot,\cdot)$ solves \eqref{eq: b fun condition}. Compared to the choice of $\psi(y,x)=|F(y,x)-1/2|$ mentioned in Section \ref{sec:theory}, we can view $\psi_*$ as having a ``shape'' adjustment $b(x,\alpha)-\alpha/2$. Since $F(Y_t,X_t)$ is independent of $X_t$, the optimal conformity score measures the distance between two independent components: $F(Y_t,X_t)$ and $1/2+(b(X_t,\alpha)-\alpha/2)$. Hence, by Lemma \ref{lem: opt score}, in order to take into account the shape  of the conditional distribution $F(\cdot,x)$, it suffices to consider the scalar quantity $1/2+(b(x,\alpha)-\alpha/2)$.

In some special cases, the ``shape'' adjustment can be shown to be zero, i.e., $b(x,\alpha)=\alpha/2$. One typical example is when $F(\cdot,x)$ is a symmetric uni-modal distribution with a well-defined conditional density.\footnote{In this case,  $Q(1/2+\delta,x)-Q(1/2,x)=Q(1/2,x)-Q(1/2-\delta,x)$ and the conditional density is increasing on $(-\infty, Q(1/2,x))$ and decreasing on $(Q(1/2,x),\infty)$. One can show $b(x,\alpha)=\alpha/2$ by taking the first-order derivative for the optimization problem in \eqref{eq: b fun condition} and setting it to zero.} Therefore, the choice of $\psi(y,x)=|F(y,x)-1/2|$ mentioned in Section \ref{sec:theory} is optimal in these cases. However, Lemma \ref{lem: opt score} provides a construction that achieves optimality more generally. By the definition of $\psi_*$ and $Q_\psi(1-\alpha)=(1-\alpha)/2$, the prediction interval is 
\begin{equation}\label{eq: interval form conformal}
\Ccal^{{\rm conf}}_{(1-\alpha)}(x)=[Q(b(x,\alpha),x), Q(b(x,\alpha)+1-\alpha,x)].
\end{equation}

We illustrate this in Figure \ref{fig:intro} with $\alpha=0.1$. \eqref{eq: interval form conformal} implies that $b(x,\alpha)$ is the quantile-index of the lower bound of the interval. For the symmetric distribution in the left panel, we see $b(x,\alpha)=0.05$, which is $\alpha/2$. For the asymmetric distribution in the right panel, we see that $b(x,\alpha)=0.007$, which is far away from $\alpha/2=0.05$.

\begin{figure}[H]
\begin{center}
\includegraphics[width=0.4\textwidth,trim=0 1.5cm 0 1cm]{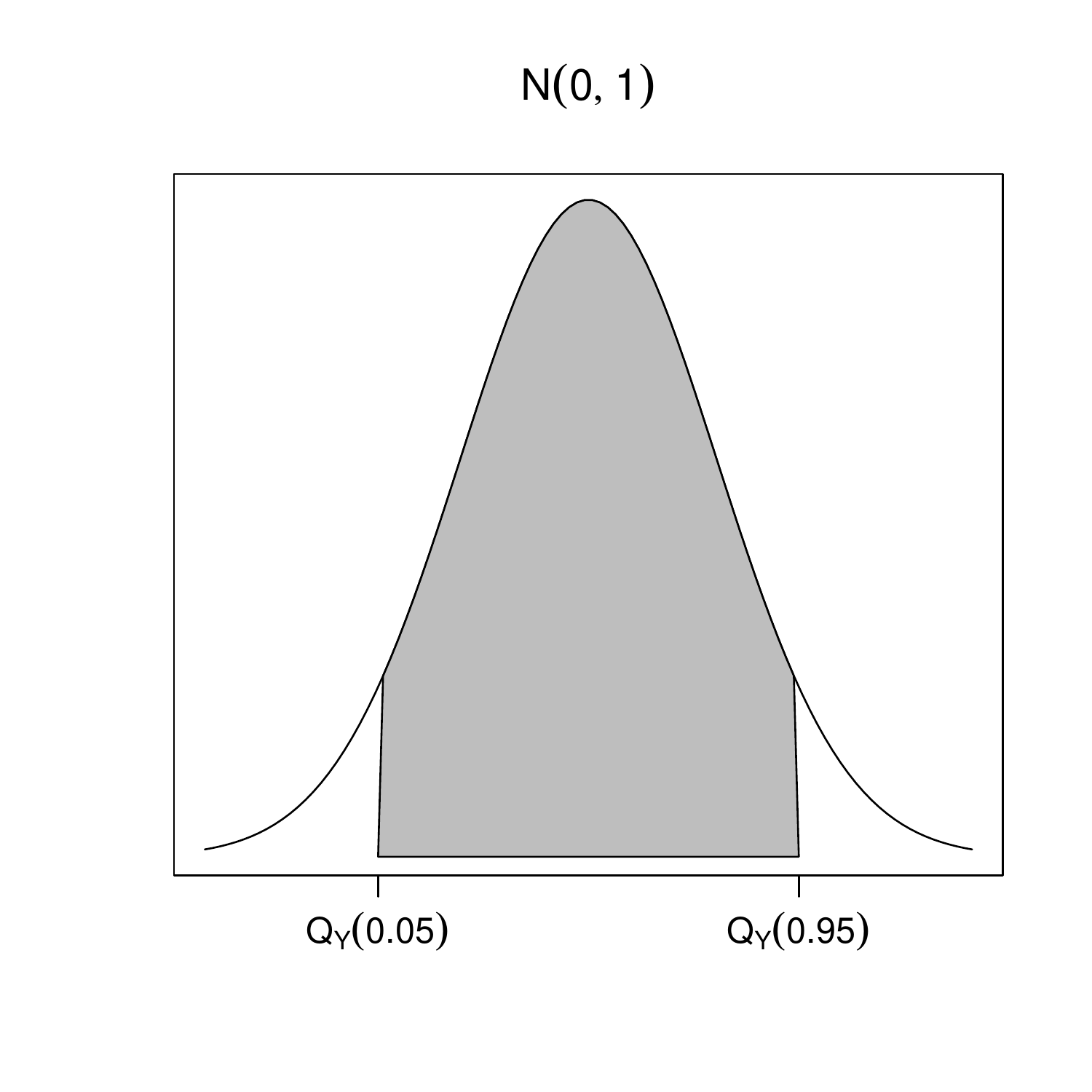}
\includegraphics[width=0.4\textwidth,trim=0 1.5cm 0 1cm]{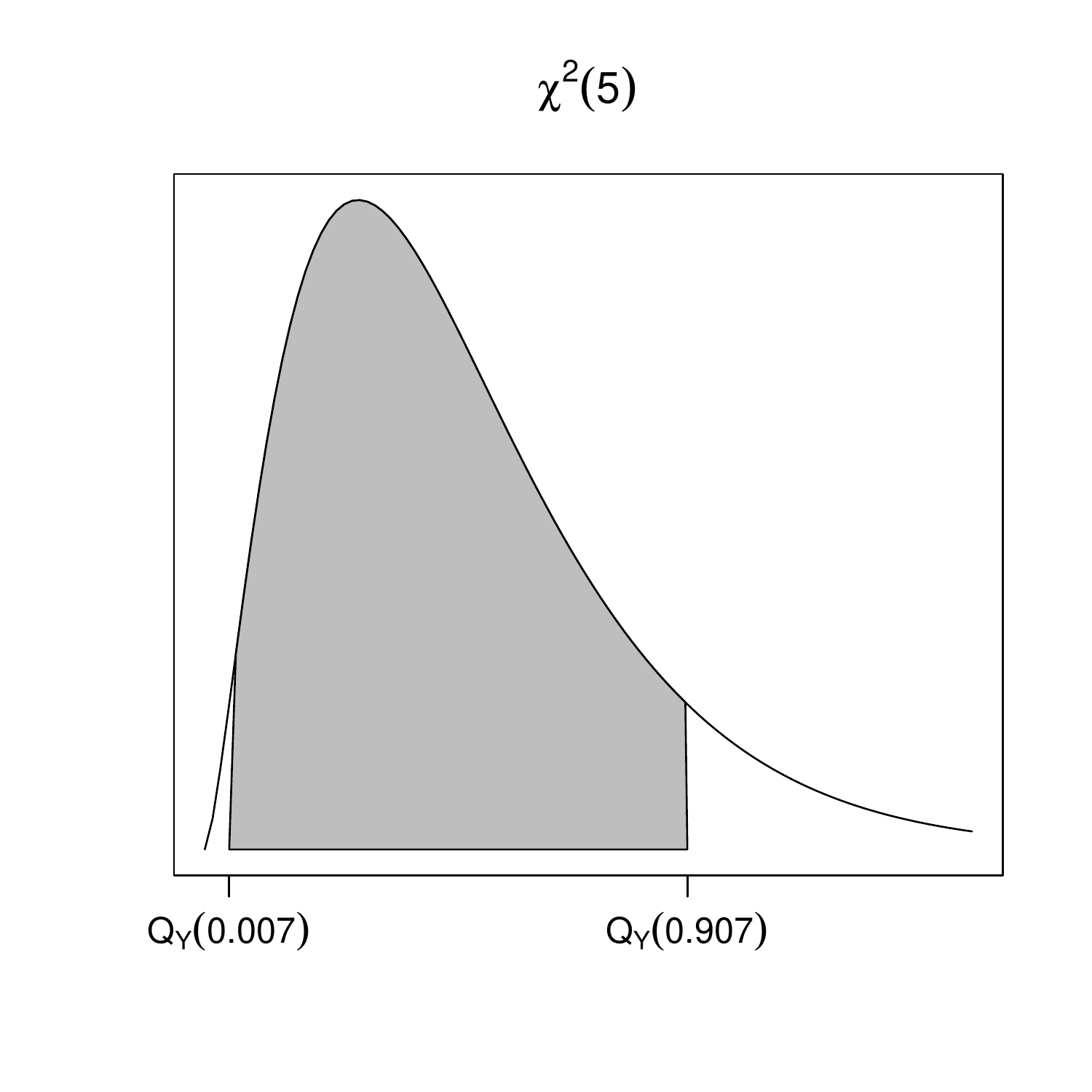}
\caption{Optimal prediction intervals}
\label{fig:intro}
\end{center}
\end{figure}

The first result in Lemma \ref{lem: opt score} is general and allows for the lack of  uniqueness of the optimal prediction interval. For example, if $F$ is the uniform distribution on a certain interval, then all conditionally valid prediction intervals have the same length. Clearly, in this case, achieving the optimal length is the only goal one can hope for.
When we can uniquely define the optimal prediction interval, Lemma \ref{lem: opt score} implies that the conformal procedure can recover the uniquely defined optimal interval, not just achieving the optimal length. 

Lemma \ref{lem: opt score} also confirms the insight of \cite{lei2014distribution}: the optimal confidence set for $X_{T+1}=x$ should take the form $ \{y:\ f(y,x)\geq c(x)\}$ for some $c(x)>0$, where $f(y,x)=\partial F(y,x)/ \partial y$. Assume that $F(\cdot,x)$ is a uni-modal distribution  and $f(\cdot,x)$ is a continuous function for any $x\in\Xcal$. Then this confidence set is an interval. This means that $ \{y:\ f(y,x)\geq c(x)\}=[c_1(x),c_2(x)]$ and $f(c_1(x),x)=f(c_2(x),x)=c(x)$. We notice that $c_1(x),c_2(x)$ are related to our results in that $c_1(x)=Q(b(x,\alpha),x)$ and $c_2(x)=Q(b(x,\alpha)+1-\alpha,x)$. To see this, simply observe that the first-order condition of the optimization problem in \eqref{eq: b fun condition} is 
$1/f(Q(z+1-\alpha,x),x)-1/f(Q(z,x))=0 $, which implies that 
$$
f(Q(b(x,\alpha)+1-\alpha,x))=f(Q(b(x,\alpha),x)).
$$

To make the procedure operational, we provide the conformal prediction interval $\widehat{\Ccal_{(1-\alpha)}^{{\rm conf}}}(X_{T+1})$ defined
in Algorithm \ref{algo: optimal} in the SI Appendix. We can provide the following guarantee. 

\begin{thm}\label{thm: effigency gen 2}
Let Assumption \ref{assu: efficiency gen} in the SI Appendix  hold.  Then 
\[
P\left(Y_{T+1}\in\widehat{\Ccal_{(1-\alpha)}^{{\rm conf}}}(X_{T+1})\mid X_{T+1}\right)=1-\alpha+o_{P}(1)
\]
and 
\[
\mu\left(\widehat{\Ccal_{(1-\alpha)}^{{\rm conf}}}(X_{T+1})\right)\leq\mu\left(\Ccal_{(1-\alpha)}^{{\rm opt}}(X_{T+1})\right)+o_{P}(1).
\]
\end{thm}

The main requirements in Assumption \ref{assu: efficiency gen} in the SI Appendix are consistency of $\hat F$ and that the density $f$ bounded below on its support. This is quite mild in the sense that it does not imply that the optimal prediction interval in \eqref{eq: opt con int Y} is uniquely defined. For example, it allows $f$ to be a uniform distribution. Therefore, as discussed above, the conformal prediction interval would have approximately the shortest length but might not converge to $\Ccal^{{\rm opt}}_{(1-\alpha)}(X_{T+1})$  in \eqref{eq: opt con int Y}. 

The following theorem provides a stronger result about $\widehat{\Ccal_{(1-\alpha)}^{{\rm conf}}}(X_{T+1})$ based on stronger assumptions.

\begin{thm}
\label{thm: efficiency}Let Assumption \ref{assu: efficiency } in the SI Appendix hold.
Consider the conformal prediction interval $\widehat{\Ccal_{(1-\alpha)}^{{\rm conf}}}(X_{T+1})$ defined
in Algorithm \ref{algo: optimal} in the SI Appendix. Then 
\[
\mu\left(\widehat{\Ccal_{(1-\alpha)}^{{\rm conf}}}(X_{T+1})\triangle\Ccal_{(1-\alpha)}^{{\rm opt}}(X_{T+1})\right)=o_{P}(1),
\]
where $\triangle$ denotes the symmetric difference of sets (i.e.,
$A\triangle B=(A\backslash B)\bigcup(B\backslash A)$), $\Ccal_{(1-\alpha)}^{{\rm opt}}(X_{T+1})$
is defined in \eqref{eq: opt con int Y}.
\end{thm}

The key component of  Assumption \ref{assu: efficiency } in the SI Appendix  is consistent estimation of $b$. Theorem \ref{thm: efficiency} shows that  $\widehat{\Ccal_{(1-\alpha)}^{{\rm conf}}}(X_{T+1})$ is close to $\Ccal_{(1-\alpha)}^{{\rm opt}}(X_{T+1})$ in the sense that the symmetric difference between these two sets has vanishing Lebesgue measure.

\section{Empirical Applications} 
\label{sec:applications}

We illustrate the performance of DCP in two empirical applications and provide a comparison to alternative approaches. We consider eight  different conformal prediction methods.

\begin{enumerate}\setlength\itemsep{0pt}
    \item \textbf{DCP-QR:} DCP with QR (Algorithm \ref{algo: split dcp})
    \item \textbf{DCP-QR$^\ast$:} Optimal DCP with QR (Algorithm \ref{algo: optimal} in SI Appendix)
    \item \textbf{DCP-DR:} DCP with DR (Algorithm \ref{algo: split dcp})
    \item \textbf{CQR:} CQR with QR \citep{romano2019conformalized}
    \item \textbf{CQR-m:} CQR variant \citep{sesia2019comparison,kivaranovic2020adaptive} with QR
    \item \textbf{CQR-r:} CQR variant \citep{sesia2019comparison} with QR.
    \item \textbf{CP-OLS:} Mean-based split conformal prediction with OLS
    \item \textbf{CP-loc:} Locally-weighted conformal prediction \citep{lei2018distributionfree} with OLS
\end{enumerate}

All computations were carried out in \texttt{R} \citep{R2021}. Code and data for replicating the empirical results are deposited on Github (\url{https://github.com/kwuthrich/Replication_DCP}).

\subsection{Predicting Stock Market Returns}
\label{sec:predicting_stock_returns}

Here we consider the problem of predicting stock market returns, which are known to exhibit substantial heteroskedasticity; see Chapter 13 in \cite{elliott2016economic} for a recent review and the references therein. We use data on daily returns of the market portfolio (CRSP value-weighted portfolio) from July 1, 1926, to June 30, 2021.\footnote{The CRSP data are constructed from the Fama/French 3 Factors data \citep{data_returns} available from \href{https://mba.tuck.dartmouth.edu/pages/faculty/ken.french/data_library.html}{Kenneth R. French's data library} (accessed August 17, 2021).} We use lagged realized volatility $X_t$ to predict the present return $Y_t$.\footnote{We compute realized volatility as the square root of the sum of squared returns over the last 22 days.} Daily returns are not iid and exhibit time series dependence. In the SI Appendix, we show that the key conditions underlying our theoretical results hold when the data are $\beta$-mixing. Several stochastic volatility models for asset returns, including the popular GARCH models, can be shown to be $\beta$-mixing \citep[e.g.,][]{boussama1998ergodicite,carrasco2002mixing,francq2006mixing}.

We evaluate the performance of the different methods by splitting the data into a holdout and a test sample. To account for the dependence in the data, we present results averaged over five consecutive prediction exercises. In the first exercise, we apply split conformal prediction with an equal split ($|\mathcal{T}_1|=|\mathcal{T}_2|$) to the first 50\% of observations and use the next 10\% for testing. In the second exercise, we drop the first 10\% of the observations, apply split conformal prediction to the next 50\% of observations, and use the next 10\% for testing and so on.

\begin{figure}[H]
\begin{center}
\includegraphics[width=0.8\textwidth,trim=0 1.5cm 0 0.5cm]{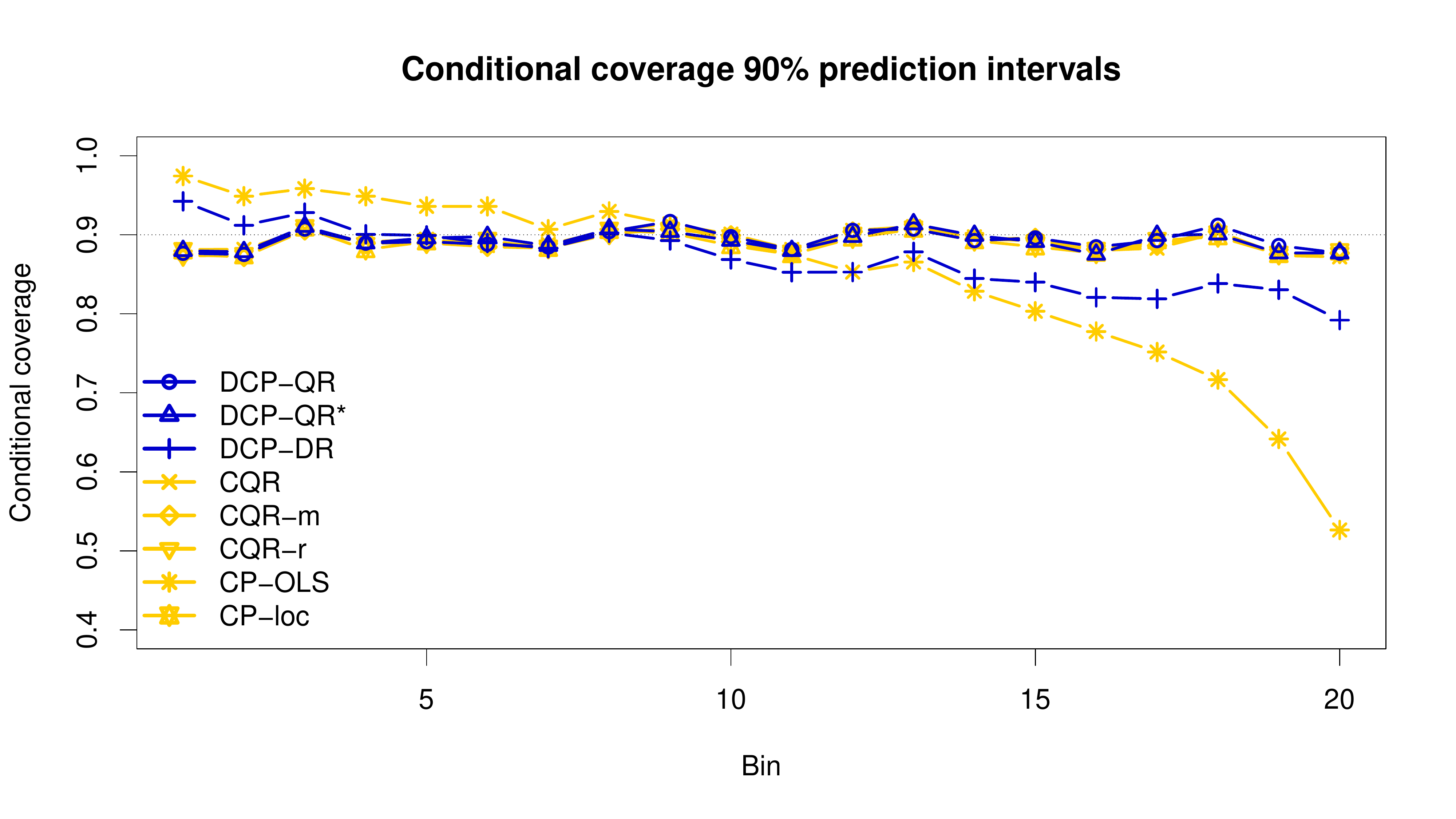}
\caption{Conditional coverage 90\% prediction intervals by realized volatility}
\label{fig:conditional_coverage_returns}
\end{center}
\end{figure}

Figure \ref{fig:conditional_coverage_returns} plots the empirical coverage probabilities for 20 bins obtained by dividing up the support of $X_t$ based on equally spaced quantiles. DCP-QR and DCP-QR$^\ast$ yield prediction intervals with coverage levels that are almost constant across all bins and close to the nominal level. They outperform DCP-DR, which undercovers in high-volatility regimes. The conditional coverage properties of DCP-QR and DCP-QR$^\ast$ are very similar to CQR, CQR-m, CQR-r, and CP-loc. This suggest that location-scale models, which are nested by QR, provide a good approximation of the conditional distribution. CP-OLS exhibits overcoverage under low-volatility regimes and substantial undercoverage under high-volatility regimes. This finding has important practical implications since the volatility tends to be high during periods of crisis, which is precisely when accurate risk assessments are most needed.

\begin{figure}[H]
\begin{center}
\includegraphics[width=0.8\textwidth,trim=0 1.5cm 0 0.5cm]{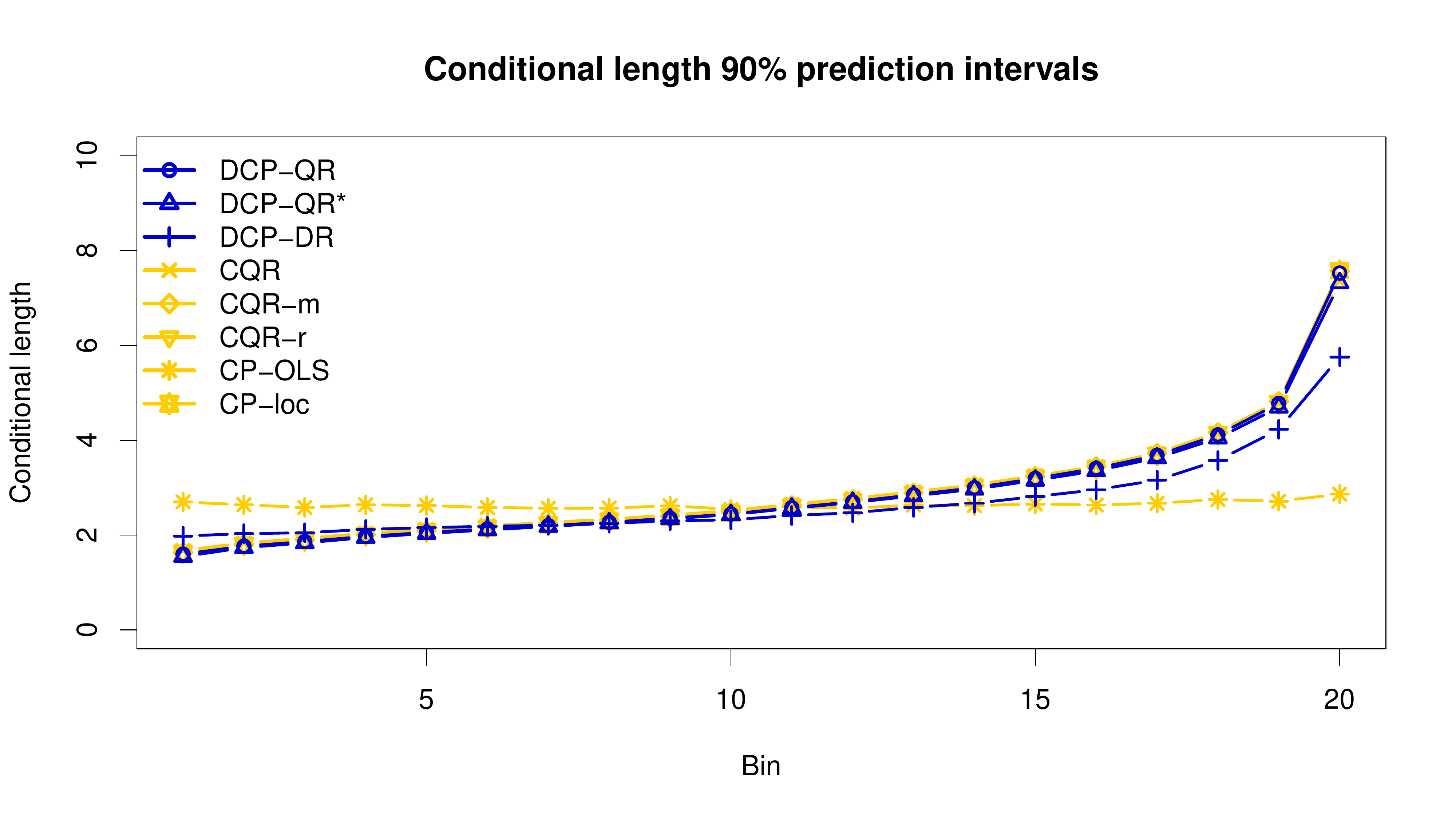}
\caption{Conditional coverage 90\% prediction intervals by realized volatility}
\label{fig:conditional_length_returns}
\end{center}
\end{figure}

Figure \ref{fig:conditional_length_returns} shows the conditional length of the prediction intervals. DCP-QR, DCP-QR$^\ast$, CQR, CQR-m, CQR-r, and CP-loc yield prediction intervals of similar length.
The DCP-DR prediction intervals are somewhat shorter than those of the QR-based methods at the upper tail. Finally, CP-OLS yields prediction intervals that are almost constant across all values of realized volatility; they are longer at the lower tail and shorter at the upper tail.\footnote{The CP-OLS prediction intervals are not exactly constant because we are reporting results averaged over five experiments.}

\subsection{Predicting Wages Using CPS Data}
\label{sec:predicting_wages}
We consider the problem of predicting wages using individual characteristics. We use the 2012 CPS data provided in the \texttt{R}-package \texttt{hdm} \citep{hdm2016}, which contains information on $N=29217$ observations. Here we use the index $i$ instead of $t$. To illustrate the impact of skewness on the performance of the different prediction methods, we use the hourly wage as our dependent variable $Y_i$.\footnote{We obtain the hourly wage by exponentiating the log hourly wage provided in the dataset.}  Predictors $X_i$ include indicators for gender, marital status, educational attainment, region, experience, experience squared, and all two-way interactions such that $\dim(X_i)=100$ after removing constant variables. 

Following \cite{romano2019conformalized} and \cite{sesia2019comparison}, we evaluate the performance of the different methods by randomly holding out 20\% of the data for testing, $\mathcal{I}_{\rm test}$, and applying split conformal  prediction with an equal split to the remaining 80\% of the data. We repeat the whole experiment 20 times. 

Panel (a) of Table \ref{tab:coverage} shows that all conformal prediction methods exhibit excellent unconditional coverage properties, confirming the theoretical finite sample guarantees. To assess and compare the conditional coverage properties, for each method, we compute conditional coverage probabilities as the predictions from logistic regressions of  $\{Y_i\in \widehat{\mathcal{C}^{\rm split}_{(1-\alpha)}}(X_i)\}_{i\in \mathcal{I}_{\rm test}}$ on $\{X_i\}_{i\in \mathcal{I}_{\rm test}}$, where  
$\widehat{\mathcal{C}^{\rm split}_{(1-\alpha)}}$ is the split conformal prediction interval obtained by the corresponding method.  The less dispersed the predicted coverage probabilities are around the nominal level $1-\alpha=0.9$, the better the overall conditional coverage properties of a method. Panel (b) of Table \ref{tab:coverage} plots the standard deviation of the predicted coverage probabilities.\footnote{Using $\sqrt{1/| \mathcal{I}_{\rm test}|\sum_{i \in \mathcal{I}_{\rm test}}(\widehat{\texttt{Coverage}}_i-0.9)^2}$, where $\widehat{\texttt{Coverage}}_i$ is the predicted coverage probability, instead of the standard deviation yields very similar results.}  DCP-QR$^\ast$  yields the lowest dispersion of all methods. The predicted coverage probabilities based on DCP-QR are less dispersed than those obtained from CQR, CQR-m, CQR-r. CP-loc yields a higher dispersion than the methods based on QR and DR, which demonstrates the value-added of using flexible models of the conditional distribution. Overall, DCP performs much better than CP-OLS for which the predicted coverage probabilities exhibit a very high dispersion. Figure \ref{fig:conditional_coverage_cps} in the SI Appendix plots histograms of the predicted coverage probabilities.

Table \ref{tab:length} shows the average length of the prediction intervals. DCP-QR$^\ast$ produces the shortest prediction intervals among of all methods. This demonstrates the practical advantage of the shape adjustment when the conditional distribution is skewed. The results also suggest a trade-off between conditional coverage accuracy and average length. For example, CP-OLS and CP-loc, which both exhibit poor conditional coverage properties, yield shorter prediction intervals than DCP-QR.

\begin{table}[ht]
\centering
\caption{Coverage 90\% prediction intervals}
\setlength{\tabcolsep}{2pt} 
\begin{tabular}{cccccccc}
\toprule
\midrule
DCP-QR&DCP-QR$^\ast$&DCP-DR&CQR&CQR-m&CQR-r&CP-OLS&CP-loc\\
\midrule
\multicolumn{8}{c}{(a) Unconditional coverage} \\
\cmidrule(l{5pt}r{5pt}){1-8}   \ 
0.90 & 0.90 & 0.90 & 0.90 & 0.90 & 0.90 & 0.90 & 0.90 \\ 
\midrule
\multicolumn{8}{c}{(b) Dispersion of predicted conditional coverage ($\times 100$)} \\
\cmidrule(l{5pt}r{5pt}){1-8}   \ 
1.80 & 1.71 & 3.08 & 2.21 & 2.36 & 2.30 & 11.13 & 4.11 \\ 
\midrule
\bottomrule
\end{tabular}
\label{tab:coverage}
\end{table}

\begin{table}[ht]
\centering
\caption{Average length 90\% prediction intervals}
\setlength{\tabcolsep}{2pt} 
\begin{tabular}{cccccccc}
\toprule
\midrule
DCP-QR&DCP-QR$^\ast$&DCP-DR&CQR&CQR-m&CQR-r&CP-OLS&CP-loc\\
\midrule
34.22 & 29.61 & 33.69 & 34.52 & 34.84 & 34.63 & 33.84 & 32.66 \\ 
\midrule
\bottomrule
\end{tabular}
\label{tab:length}
\end{table}

\bibliographystyle{apalike}
\bibliography{pnas-biblio}

\newpage

\appendix
\begin{center}
\LARGE{SI Appendix}
\end{center}

\section{Details for  Section \ref{sec:extension}}
\label{sec:optimal_interval}
Here we describe in detail how to implement the optimal prediction intervals described in Section \ref{sec:extension}.

\subsection{Implementation}

We now consider the estimation of $b(\cdot,\cdot)$. We assume that $\hF(y,x)$ is monotonic
in $y$; if not, we first rearrange it. We define $\hQ(\cdot,\cdot)$ by 
\[
\hQ(\tau,x)=\inf\left\{ y:\ \hF(y,x)\geq\tau\right\} .
\]

Define 
\[
L(x)=\min_{z\in[0,\alpha]}Q(z+1-\alpha,x)-Q(z,x)
\]
and 
\[
\hL(x)=\min_{z\in[0,\alpha]}\hQ(z+1-\alpha,x)-\hQ(z,x).
\]

Let $\hb(x,\alpha)$ be a function such that $\hb(x,\alpha)\in[0,\alpha]$ and
\begin{equation}
\hQ(\hb(x,\alpha)+1-\alpha,x)-\hQ(\hb(x,\alpha),x)=\hL(x). \label{eq: b estimator sample}
\end{equation}

We propose the following algorithm.

\begin{customalgo}{S1}[Optimal split DCP]
\label{algo: optimal}
\text{ }

\begin{itemize}\setlength{\itemsep}{0pt}\setlength{\parskip}{0pt}
\item[]\textbf{Input:}  Data $\{(Y_{t},X_{t})\}_{t=1}^{T}$, miscoverage level $\alpha\in(0,1)$,
and point $X_{T+1}$
\item[] \textbf{Process:}
\begin{enumerate}
\item Split $\{1,\dots,T+1\}$ into $\mathcal{T}_{1}:=\{1,\dots,T_{0}\}$ and
$\mathcal{T}_{2}:=\{T_{0}+1,\dots,T\}$.
\item Obtain $\hF$ and $\hb$ based on $\{Z_{t}\}_{t\in\mathcal{T}_{1}}$.
\item Compute $\{\hV_{t}^{*}\}_{t\in\mathcal{T}_{2}}$ with $\hV_{t}^{*}=\hF(Y_{t},X_{t})-\hb(X_{t},\alpha)-\frac{1}{2}(1-\alpha)$.
\item Compute $\hQ_{\mathcal{T}_{2}}^{*}$, the $(1-\alpha)(1+1/|\mathcal{T}_{2}|)$
empirical quantile of $\left\{ |\hV_{t}^{*}|\right\} _{t\in\mathcal{T}_{2}}$.
\end{enumerate}
\item[] \textbf{Output:}  Return the prediction set
\[
\widehat{\Ccal_{(1-\alpha)}^{{\rm conf}}}(X_{T+1})=\left\{ y:\ \left|\hF(y,X_{t})-\hb(X_{T+1},\alpha)-\frac{1}{2}(1-\alpha)\right|\leq\hQ_{\mathcal{T}_{2}}^{*}\right\} .
\]
(Since $\hF(\cdot,X_{T+1})$ is monotonic, $\widehat{\Ccal_{(1-\alpha)}^{{\rm conf}}}(X_{T+1})$
is an interval.)
\end{itemize}
\end{customalgo}

\subsection{Regularity conditions}

We now provide the regularity condition for Theorem \ref{thm: effigency gen 2}. For simplicity we focus on the case of iid data. Recall that a legitimate cumulative distribution function $F(\cdot)$
on $\RR$ is a non-decreasing right-continuous function such that
$\lim_{z\rightarrow-\infty}F(z)=0$ and $\lim_{z\rightarrow\infty}F(z)=1$. 

\begin{customass}{S1}
\label{assu: efficiency gen}Suppose that the following 
hold:
\begin{enumerate}
\item The data $\{(Y_{t},X_{t})\}_{t\in\Tcal_{2}}$ is iid and $|\Tcal_{2}|\rightarrow\infty$. 
\item For any $x\in\Xcal$, $\hF(\cdot,x)$ is a legitimate cumulative distribution
function on $\RR$ with probability one.
\item $\sup_{x\in\Xcal}\sup_{y\in\mathcal{Y}(x)}|\hF(y,x)-F(y,x)|=o_{P}(1)$
as $|\Tcal_{1}|\rightarrow\infty$, where $\mathcal{Y}(x)$ is the
support of conditional distribution $Y_{t}\mid X_{t}=x$. 
\item There exist constants $C_{1},C_{2}>0$ such that $\min_{y\in\mathcal{Y}(x)}f(y,x)\geq C_{1}$
and $\sup_{y\in\mathcal{Y}(x)}|y|\leq C_{2}$ for any $x\in\Xcal$.
\end{enumerate}
 \end{customass}

The following assumption is used to prove the results in Theorem  \ref{thm: efficiency}.

\begin{customass}{S2}
\label{assu: efficiency }Suppose that the following hold. 
\begin{enumerate}
\item $|\mathcal{T}_{2}|^{-1}\sum_{t\in\mathcal{T}_{2}}(\hF(Y_{t},X_{t})-U_{t})^{2}=o_{P}(1)$
and $|\mathcal{T}_{2}|^{-1}\sum_{t\in\mathcal{T}_{2}}(\hb(X_{t},\alpha)-b(X_{t},\alpha))^{2}=o_{P}(1)$, where $b(\cdot,\cdot)$ is the unique function satisfying the requirement in Lemma \ref{lem: opt score}.
\item $\sup_{v\in\RR}|\tilde{G}_{*}(v)-G_{*}(v)|=o_{P}(1)$, where $\tG_{*}(v)=|\mathcal{T}_{2}|^{-1}\sum_{t\in\mathcal{T}_{2}}\oneb\{\hV_{t}^{*}\leq v\}$
and $G_{*}(\cdot)$ is the distribution function of $V_{t}^{*}=U_{t}-b(X_{t},\alpha)-\frac{1}{2}(1-\alpha)$. 
\item There exists a constant $C_{1}>0$ such that for any $x\in\Xcal$,
$\inf_{y\in s(x)}f(y,x)\geq C_{1}$, where $s(x)=[s_{1}(x),s_{2}(x)]$
is the support of the distribution $Y\mid X=x$.
\item $\sup_{y\in\RR}\left|\hF(y,X_{T+1})-F(y,X_{T+1})\right|=o_{P}(1)$
and $\hb(X_{T+1},\alpha)=b(X_{T+1},\alpha)+o_{P}(1)$.
\item There exists a constant $C_2>0$ such that for any $x\in\Xcal$, $\max\{|s_1(x)|,|s_2(x)|\}\leq C_2$.
\end{enumerate}
 \end{customass}

The key requirement in Assumption \ref{assu: efficiency } is the consistency of $\hb$. Since $\hb$ is a solution to the optimization problem in \eqref{eq: b estimator sample}, we can establish its consistency using the same argument for  the consistency of an M-estimator. Under Assumption \ref{assu: efficiency gen}, we only need to impose the convexity  of the mapping $z\mapsto Q(z+1-\alpha,x)-Q(z)$. A  simple sufficient condition is that  there exists constants $\kappa_1,\kappa_2>0$ such that for any $x\in\Xcal$ and for any $z$ with $|b(x,\alpha)-z|\leq \kappa_1$, 
$$
\frac{\partial^2}{\partial z^2} \left(Q(z+1-\alpha,x)-Q(z) \right) \geq \kappa_2.
$$

For uni-modal distributions, the above condition can be verified once we assume that the density $f(\cdot,x)$ is not too flat around $Q(b(x,\alpha),x)$ and  $Q(b(x,\alpha)+1-\alpha,x)$. A similar condition is imposed as Assumption 2 in \cite{lei2014distribution}.

\section{Regression models for conditional distributions}
\label{sec:models}
An important advantage of the proposed approach is that it allows researchers to leverage powerful regression methods for estimating conditional CDFs. This section discusses semiparametric (and potentially penalized) QR and DR models, which are very popular in applied research. We emphasize that our method is generic and also works in conjunction with nonparametric estimators \citep[e.g.,][]{chaudhuri1991,koenker1994splines,he1998bivariate} as well as high-dimensional methods based on trees and random forests \citep[e.g.,][]{chaudhuri2002,meinshausen2006quantile} and neural networks \citep[e.g.,][]{taylor2000quantile}.

\subsection{Quantile regression methods}

QR methods impose a model for the conditional quantiles $Q(\tau, x)$. The implied model for the conditional CDF is \citep[][]{chernozhukov2013inference}
\begin{equation}
F(y,x)=\int_{0}^1\oneb\left\{ Q(\tau, x)\le y\right\}d\tau.
\end{equation}
A leading example is where $Q(\tau, x)$ is assumed to be linear:
\begin{equation}
Q\left(\tau,x\right)=x^\top\beta(\tau)
\end{equation}
If $X_t$ is low dimensional, the parameter of interest $\beta(\tau)$ can be estimated using linear QR \citep{koenker1978} as the solution to a convex program
\begin{equation}
\hat\beta(\tau)\in \arg\min_{\beta \in \mathbb{R}^p}\sum_{t=1}^{T+1}\rho_\tau \left( Y_t-X_t^\top\beta \right), \label{eq:qr}
\end{equation}
where $\rho_\tau(u) := u(\tau-\oneb\{u<0\})$ is the check function. In problems where $X_t$ is high-dimensional, it may be convenient to consider a penalized version of  program \eqref{eq:qr}:
\begin{equation}
\hat\beta(\tau)\in \arg\min_{\beta \in \mathbb{R}^p}\sum_{t=1}^{T+1}\rho_\tau \left( Y_t-X_t^\top\beta \right)+\mathcal{P}(\beta),
\end{equation}
where $\mathcal{P}(\beta)$ is a penalty function. Examples of $\mathcal{P}(\beta)$ include $\ell_1$-penalties \citep[e.g.,][]{koenker2004,li2008ell1,belloni2011L1} and SCAD \citep[e.g.,][]{wu2009variable}. The conditional distribution can be estimated as
\[
\hat{F}(y ,x)=\int_0^1 \oneb\left\{x^\top\hat\beta(\tau)\le y\right\}d\tau.
\]

\subsection{Distribution regression methods}

Instead of modeling the conditional quantile function, one can directly model the conditional CDF using DR \citep[e.g.,][]{foresi1995conditional,chernozhukov2013inference,chernozhukov2020generic}. DR methods impose a generalized linear model for the CDF:
\[
F(y, x)=\Lambda \left(x^\top\beta(y) \right),
\]
where $\beta(y)$ is the parameter of interest and $\Lambda(\cdot)$ is a known link function, for example, the Probit or Logit link.

If $X_t$ is low dimensional, the parameters $\beta(y)$ can be estimated as
\begin{equation}
\hat\beta(y)\in \arg\max_{\beta \in \mathbb{R}^p}\sum_{t=1}^{T+1}\left[\oneb\left\{Y_t\le y \right\} \log\left(\Lambda \left(X_t^\top\beta \right)\right)+ \oneb\left\{Y_t> y \right\} \log\left(1-\Lambda \left(X_t^\top\beta \right)\right)\right]. \label{eq:dr}
\end{equation}
When $\Lambda(\cdot)$ is the Probit (Logit) link, this is simply a Probit (Logit) regression of $1\left\{Y_t\le y \right\}$ on $X_t$. In high dimensional settings, one can use a penalized version of program \eqref{eq:dr} (e.g., with an $\ell_1$-penalty or an elastic net penalty). The conditional distribution can be estimated as $\hat{F}(y, x)=\Lambda \left(x^\top\hat\beta(y) \right)$.

\section{Proofs}

\subsection{Proof of Lemma \ref{lem: basic}}

	The argument is similar to Theorem 2 in \cite{chern2018COLT}. Let $\delta>0$ be a 
	constant to be chosen later. Define quantities $R_{T}=\sup_{v\in\RR}|\tG(v)-G(v)|$
	and $W=\sup_{x_{1}\neq x_{2}}|G(x_{1})-G(x_{2})|/|x_{1}-x_{2}|$. 
	
	Let $A=\{t:\ |\hV_{t}-V_{t}|\geq\delta\}$. Fix $x\in\RR$. Then 
	\begin{align}
	& (T+1)\left|\hG(x)-\tG(x)\right|\nonumber \\
	& \leq\left|\sum_{t\in A}\left(\oneb\{\hV_{t}< x\}-\oneb\{V_{t}< x\}\right)\right|+\left|\sum_{t\in A^{c}}\left(\oneb\{\hV_{t}< x\}-\oneb\{V_{t}<x\}\right)\right|\nonumber \\
	& \overset{\text{(i)}}{\leq}|A|+\left|\left(\sum_{t\in A^{c}}\oneb\{\hV_{t}< x\}\right)-\left(\sum_{t\in A^{c}}\oneb\{V_{t}< x\}\right)\right|\label{eq: basic lem 1}
	\end{align}
	where (i) follows by the fact that the difference of two indicators takes value in $\{-1,0,1\}$. We notice
	that for $t\in A^{c}$, $V_{t}-\delta<\hV_{t}< V_{t}+\delta$.
	Therefore, 
	\[
	\sum_{t\in A^{c}}\oneb\{V_{t}< x-\delta\}\leq\sum_{t\in A^{c}}\oneb\{\hV_{t}< x\}\leq\sum_{t\in A^{c}}\oneb\{V_{t}< x+\delta\}.
	\]
	Since $\sum_{t\in A^{c}}\oneb\{V_{t}\leq x\}$ is also between $\sum_{t\in A^{c}}\oneb\{V_{t}\leq x-\delta\}$
	and $\sum_{t\in A^{c}}\oneb\{V_{t}\leq x+\delta\}$, it follows that
	\begin{align*}
	& \left|\left(\sum_{t\in A^{c}}\oneb\{\hV_{t}< x\}\right)-\left(\sum_{t\in A^{c}}\oneb\{V_{t}< x\}\right)\right|\\
	& \leq\left|\left(\sum_{t\in A^{c}}\oneb\{V_{t}\leq x+\delta\}\right)-\left(\sum_{t\in A^{c}}\oneb\{V_{t}\leq x-\delta\}\right)\right|\\
	& =\left|(T+1)\left[\tG(x+\delta)-\tG(x-\delta)\right]-\left(\sum_{t\in A}\oneb\{V_{t}\leq x+\delta\}\right)+\left(\sum_{t\in A}\oneb\{V_{t}\leq x-\delta\}\right)\right|\\
	& \leq(T+1)\left[\tG(x+\delta)-\tG(x-\delta)\right]+\left|\sum_{t\in A}\left(\oneb\{V_{t}\leq x+\delta\}-\oneb\{V_{t}\leq x-\delta\}\right)\right|\\
	&  \overset{\text{(i)}}{\leq}(T+1)\left(G(x+\delta)-G(x-\delta)+2R_{T}\right)+|A|\\
	& \leq(T+1)\left(2\delta W+2R_{T}\right)+|A|,
	\end{align*}
	where (i) follows by the fact that the difference of two indicators takes value in $\{-1,0,1\}$. Combining the above display with (\ref{eq: basic lem 1}), we obtain
	that 
	\[
	(T+1)\left|\hG(x)-\tG(x)\right|\leq 2|A|+(T+1)\left(2\delta W+2R_{T}\right).
	\]
	
	Since the right-hand side does not depend on $x$, we have that 
	\[
	\sup_{x\in\RR}\left|\hG(x)-\tG(x)\right|\leq 2\frac{|A|}{T+1}+2\delta W+2R_{T}.
	\]
	
	To bound $|A|$, we notice that 
	\[
	|A|\phi(\delta)\leq\sum_{t\in A}\phi(|\hV_{t}-V_{t}|)\leq\sum_{t=1}^{T+1}\phi(|\hV_{t}-V_{t}|)\leq o_{P}(T+1).
	\]
	
	Hence, the above two displays imply that 
	\begin{equation}
	\sup_{x\in\RR}\left|\hG(x)-G(x)\right|\leq\sup_{x\in\RR}\left|\hG(x)-\tG(x)\right|+R_{T}\leq o_{P}(1/\phi(\delta))+2\delta W+3R_{T}.\label{eq: basic lem 2}
	\end{equation}
	
	Now we fix an arbitrary $\eta\in(0,1)$. Choose $\delta=\eta/(6W)$.
	Since $1/\phi(\delta)$ is a constant and $R_{T}=o_{P}(1)$ by assumption,
	(\ref{eq: basic lem 2}) implies that 
	\begin{align*}
	& \limsup_{T\rightarrow\infty}P\left(\sup_{x\in\RR}\left|\hG(x)-\tG(x)\right|>\eta\right)\\
	& \leq\limsup_{T\rightarrow\infty}P\left(|o_{P}(1/\phi(\delta))|>\eta/3\right)+\limsup_{T\rightarrow\infty}P\left(|2\delta W|>\eta/3\right)+\limsup_{T\rightarrow\infty}P\left(|R_{T}|>\eta/9\right)=0.
	\end{align*}
	
	Since $\eta>0$ is arbitrary, we have 
	\[
	\sup_{x\in\RR}\left|\hG(x)-G(x)\right|=o_{P}(1).
	\]
	
	Thus, 
	\[
	\hG(\hV_{T+1})=G(\hV_{T+1})+o_{P}(1)\overset{\text{(i)}}{=}G(V_{T+1})+o_{P}(1),
	\]
	where (i) follows by $|G(\hV_{T+1})-G(V_{T+1})|\leq W|\hV_{T+1}-V_{T+1}|$.
	The proof is complete.

\subsection{Proof of Theorem \ref{thm: unconditional validity asy}}
Notice that 
$$
P\left(Y_{T+1}\in\widehat{\mathcal{C}^{\rm full}_{(1-\alpha)}}\left(X_{T+1} \right)\right)=P\left(\frac{1}{T+1}\sum_{t=1}^{T+1}\oneb\left\{\hV^{(Y_{T+1})}_{t}\ge \hV^{(Y_{T+1})}_{T+1} \right\}>\alpha\right)=P\left(1-\hG(\hV_{T+1})>\alpha \right).
$$
By Lemma \ref{lem: basic}, $ \hG(\hV_{T+1})=G(V_{T+1})+o_{P}(1)$. Since $G(\cdot)$ is continuous, $G(V_{T+1})$ has the uniform distribution on (0,1). The desired result follows.

\subsection{Proof of Theorem \ref{thm: conditional validity asy}}
	Notice that $U_{t}=F(Y_t,X_t)$ is independent of $X_{t}$. Since $V_{T+1}=\psi(U_{T+1})$,
	$V_{T+1}$ is also independent of $X_{T+1}$. This means that 
	\[
	P(G(V_{T+1})\leq\alpha\mid X_{T+1})=P(G(V_{T+1})\leq\alpha).
	\]
	Since $G(\cdot)$ is the distribution function of $V_{T+1}$ and is a continuous function, we have
	that $P(G(V_{T+1})\leq\alpha)=\alpha$. The desired result follows
	by Lemma \ref{lem: basic} and 
	
	\begin{footnotesize}
	$$
P\left(Y_{T+1}\in\widehat{\mathcal{C}^{\rm full}_{(1-\alpha)}}\left(X_{T+1} \right)\mid X_{T+1}\right)=P\left(\frac{1}{T+1}\sum_{t=1}^{T+1}\oneb\left\{\hV^{(Y_{T+1})}_{t}\ge \hV^{(Y_{T+1})}_{T+1} \right\}>\alpha\mid X_{T+1}\right)=P\left(1-\hG(\hV_{T+1})>\alpha \mid X_{T+1}\right).
$$
\end{footnotesize}

\subsection{Proof of Lemma \ref{lem: opt score}}

We proceed in three steps.

\textbf{Step 1:} show $Q_{\psi}(1-\alpha)=(1-\alpha)/2$. 

By the same argument as in Lemma \ref{lem: eff gen part 1} (proved later), 
\[
P\left(|F(Y_{t},X_{t})-b(X_{t},\alpha)-(1-\alpha)/2|\leq\frac{1-\alpha}{2}\mid X_{t}\right)=1-\alpha.
\]

Therefore, 
\[
P\left(|F(Y_{t},X_{t})-b(X_{t},\alpha)-(1-\alpha)/2|\leq\frac{1-\alpha}{2}\right)=1-\alpha.
\]

In other words, $Q_{\psi}(1-\alpha)=(1-\alpha)/2$. 

\textbf{Step 2:} show $\mu\left(\Ccal_{(1-\alpha)}^{{\rm opt}}(X_{T+1})\right)=\mu\left(\Ccal_{(1-\alpha)}^{{\rm conf}}(X_{T+1})\right)$. 

By the definition of $\Ccal_{(1-\alpha)}^{{\rm opt}}(X_{T+1})$, 
\begin{equation}
\mu\left(\Ccal_{(1-\alpha)}^{{\rm opt}}(X_{T+1})\right)=\min_{F(z_{2},X_{T+1})-F(z_{1},X_{T+1})\geq1-\alpha}z_{2}-z_{1}.\label{eq: lem opt score eq 4}
\end{equation}

Since $F(\cdot,X_{T+1})$ is a continuous function, we have that 
\begin{equation}
\min_{F(z_{2},X_{T+1})-F(z_{1},X_{T+1})\geq1-\alpha}z_{2}-z_{1}=\min_{F(z_{2},X_{T+1})-F(z_{1},X_{T+1})=1-\alpha}z_{2}-z_{1}.\label{eq: lem opt score eq 5}
\end{equation}

We can see this by contradiction.  Let $(z_{1}^{*},z_{2}^{*})$ be the solution to the optimization
in \eqref{eq: lem opt score eq 4}. Suppose that $F(z_{2}^{*},X_{T+1})-F(z_{1}^{*},X_{T+1})>1-\alpha$.
Notice that the mapping $g(z)=F(z,X_{T+1})-F(z_{1}^{*},X_{T+1})$
is continuous in $z$. Since $g(z_{2}^{*})>1-\alpha$ and $g(z_{1}^{*})=0<1-\alpha$.
By the intermediate value theorem, there exists $z_{2}^{**}\in(z_{1}^{*},z_{2}^{*})$
such that $g(z_{2}^{**})=1-\alpha$. Thus, $F(z_{2}^{**},X_{T+1})-F(z_{1}^{*},X_{T+1})\geq1-\alpha$
and $z_{2}^{**}-z_{1}^{*}<z_{2}^{*}-z_{1}^{*}$, contradicting the
assumption that $(z_{1}^{*},z_{2}^{*})$ is the solution to the optimization
in \eqref{eq: lem opt score eq 4}. Therefore, $F(z_{2}^{*},X_{T+1})-F(z_{1}^{*},X_{T+1})=1-\alpha$.
Therefore, we have that 
\[
\mu\left(\Ccal_{(1-\alpha)}^{{\rm opt}}(X_{T+1})\right)=\min_{F(z_{2},X_{T+1})-F(z_{1},X_{T+1})=1-\alpha}z_{2}-z_{1}.
\]

Since $F(z_{2},X_{T+1})-F(z_{1},X_{T+1})=1-\alpha$, we can write
$F(z_{2},X_{T+1})=F(z_{1},X_{T+1})+1-\alpha$, which means $z_{2}=Q(F(z_{1},X_{T+1})+1-\alpha,X_{T+1})$.
Since $F(z_{1},X_{T+1})+1-\alpha\leq1$, we have $F(z_{1},X_{T+1})\leq\alpha$.
Therefore, 
\begin{align}
\mu\left(\Ccal_{(1-\alpha)}^{{\rm opt}}(X_{T+1})\right) & =\min_{F(z_{1},X_{T+1})\in[0,\alpha]}Q(F(z_{1},X_{T+1})+1-\alpha,X_{T+1})-z_{1}\nonumber \\
 & \overset{\text{(i)}}{=}\min_{w\in[0,\alpha]}Q(w+1-\alpha,X_{T+1})-Q(w,X_{T+1}),\label{eq: lem opt score eq 6}
\end{align}
where (i) follows by a change of variables $w=F(z_{1},X_{T+1})$ (and
thus $z_{1}=Q(w,X_{T+1})$). 

We notice that 
\begin{align}
\Ccal_{(1-\alpha)}^{{\rm conf}}(X_{T+1}) & =\left\{ y:\ |F(y,X_{T+1})-b(X_{T+1},\alpha)-(1-\alpha)/2|\leq Q_{\psi}(1-\alpha)\right\} \nonumber \\
 & \overset{\text{(i)}}{=}\left\{ y:\ |F(y,X_{T+1})-b(X_{T+1},\alpha)-(1-\alpha)/2|\leq(1-\alpha)/2\right\} \nonumber \\
 & =\left\{ y:\ b(X_{T+1},\alpha)\leq F(y,X_{T+1})\leq b(X_{T+1},\alpha)+1-\alpha\right\} \nonumber \\
 & =\left[Q(b(X_{T+1},\alpha),X_{T+1}),\ \ Q(b(X_{T+1},\alpha)+1-\alpha,X_{T+1})\right],\label{eq: lem opt score eq 7}
\end{align}
where (i) follows by $Q_{\psi}=(1-\alpha)/2$. Thus, 
\begin{align*}
\mu\left(\Ccal_{(1-\alpha)}^{{\rm conf}}(X_{T+1})\right) & =Q(b(X_{T+1},\alpha)+1-\alpha,X_{T+1})-Q(b(X_{T+1},\alpha),X_{T+1})\\
 & \overset{\text{(i)}}{=}\min_{w\in[0,\alpha]}Q(w+1-\alpha,X_{T+1})-Q(w,X_{T+1}),
\end{align*}
where (i) follows by the assumption that $b(x,\alpha)\in\arg\min_{w\in[0,\alpha]}Q(w+1-\alpha,x)-Q(w,x)$
for any $x\in\Xcal$. By \eqref{eq: lem opt score eq 6}, $\mu\left(\Ccal_{(1-\alpha)}^{{\rm opt}}(X_{T+1})\right)=\mu\left(\Ccal_{(1-\alpha)}^{{\rm conf}}(X_{T+1})\right)$. 

\textbf{Step 3:} show that if $\Ccal_{(1-\alpha)}^{{\rm opt}}(x)$
is uniquely defined for any $x\in\Xcal$, then $\Ccal_{(1-\alpha)}^{{\rm opt}}(X_{T+1})=\Ccal_{(1-\alpha)}^{{\rm conf}}(X_{T+1})$.

Notice that $\Ccal_{(1-\alpha)}^{{\rm opt}}(x)=[r_{1}(x,\alpha),\ r_{2}(x,\alpha)]$,
where the pair $(r_{1}(x,\alpha),r_{2}(x,\alpha))$ uniquely solves
\[
\min_{z_{1},z_{2}}\ z_{2}-z_{1}\quad s.t.\quad F(z_{2},x)-F(z_{1},x)\geq1-\alpha.
\]

By the argument in \eqref{eq: lem opt score eq 5}, the pair $(r_{1}(x,\alpha),r_{2}(x,\alpha))$
uniquely solves 
\[
\min_{z_{1},z_{2}}\ z_{2}-z_{1}\quad s.t.\quad F(z_{2},x)-F(z_{1},x)=1-\alpha.
\]

By the same change of variables in \eqref{eq: lem opt score eq 6},
$r_{1}(x,\alpha)$ uniquely solves 
\[
\min_{z_{1}}\ Q(F(z_{1},x)+1-\alpha,x)-z_{1}\quad s.t.\quad F(z_{1},X_{T+1})\leq\alpha
\]
and $r_{2}(x,\alpha)=Q(F(r_{1}(x,\alpha),x)+1-\alpha,x)$. Similar
to \eqref{eq: lem opt score eq 6}, this can be rewritten as an optimization
problem on $[0,\alpha]$. Since $b(x,\alpha)$ solves $\min_{w\in[0,\alpha]}Q(w+1-\alpha,x)-Q(w,x)$,
we have 
\[
r_{1}(x,\alpha)=Q(b(x,\alpha),x)
\]
and $r_{2}(x,\alpha)=Q(b(x,\alpha)+1-\alpha,x)$. Thus, $\Ccal_{(1-\alpha)}^{{\rm opt}}(x)=[r_{1}(x,\alpha),\ r_{2}(x,\alpha)]=[Q(b(x,\alpha),x),\ Q(b(x,\alpha)+1-\alpha,x)]$.
By the same argument as in \eqref{eq: lem opt score eq 7}, $\Ccal_{(1-\alpha)}^{{\rm conf}}(x)=[Q(b(x,\alpha),x),\ Q(b(x,\alpha)+1-\alpha,x)]$.
Therefore, $\Ccal_{(1-\alpha)}^{{\rm opt}}(x)=\Ccal_{(1-\alpha)}^{{\rm conf}}(x)$.
Since this holds for any $x\in\Xcal$, we have completed the proof.

\subsection{Proof of Theorem \ref{thm: effigency gen 2}}

We first prove three auxiliary lemmas.

\begin{customlem}{S1}
\label{lem: eff gen part 1}Let Assumption \ref{assu: efficiency gen}
hold. Let $\tV_{t}^{*}=F(Y_{t},X_{t})-\hb(X_{t})-(1-\alpha)/2$ for
$t\in\Tcal_{2}$. Then 
\[
P\left(|\tV_{t}^{*}|\leq\frac{1-\alpha}{2}\mid X_{t}\right)=1-\alpha.
\]

Moreover, for any non-random $\delta\in[-\alpha,\alpha]$,
\[
P\left(|\tV_{t}^{*}|\leq\frac{1-\alpha}{2}+\delta\right)-(1-\alpha)\leq\delta\qquad{\rm if}\ \delta\in[-\alpha,0]
\]
and 
\[
P\left(|\tV_{t}^{*}|\leq\frac{1-\alpha}{2}+\delta\right)-(1-\alpha)\geq\delta\qquad{\rm if}\ \delta\in[0,\alpha].
\]
\end{customlem}

\begin{proof}
We show the two claims in two steps.

\textbf{Step 1:} show the first claim. 

We observe that 
\begin{align*}
P\left(|\tV_{t}^{*}|\leq\frac{1-\alpha}{2}\mid X_{t}\right) & =P\left(|U_{t}-\hb(X_{t})-(1-\alpha)/2|\leq\frac{1-\alpha}{2}\mid X_{t}\right)\\
 & =P\left(\hb(X_{t})\leq U_{t}\leq\hb(X_{t})+1-\alpha\mid X_{t}\right).
\end{align*}

Recall that $U_{t}=F(Y_{t},X_{t})$ is independent of $X_{t}$ and
has the uniform distribution on {[}0,1{]}. Since $t\in\Tcal_{2}$,
$(U_{t},X_{t})$ is independent of $\hb(\cdot)$. Since $\hb(X_{t})\in[0,\alpha]$,
we have that $[\hb(X_{t}),\hb(X_{t})+1-\alpha]\subseteq[0,1]$. Therefore,
\begin{align*}
P\left(|\tV_{t}^{*}|\leq\frac{1-\alpha}{2}\mid X_{t}\right) & =P\left(\hb(X_{t})\leq U_{t}\leq\hb(X_{t})+1-\alpha\mid X_{t}\right)\\
 & =\left(\hb(X_{t})+1-\alpha\right)-\hb(X_{t})=1-\alpha.
\end{align*}

\textbf{Step 2:} show the second claim. 

By the same argument as in Step 1, we have 
\begin{align*}
P\left(|\tV_{t}^{*}|\leq\frac{1-\alpha}{2}+\delta\mid X_{t}\right) & =P\left(|U_{t}-\hb(X_{t})-(1-\alpha)/2|\leq\frac{1-\alpha}{2}+\delta\mid X_{t}\right)\\
 & =P\left(\hb(X_{t})-\delta\leq U_{t}\leq\hb(X_{t})+1-\alpha+\delta\mid X_{t}\right)\\
 & \overset{\text{(i)}}{=}P\left(\max\{\hb(X_{t})-\delta,0\}\leq U_{t}\leq\min\{\hb(X_{t})+1-\alpha+\delta,1\}\mid X_{t}\right)\\
 & \overset{\text{(ii)}}{=}\min\{\hb(X_{t})+1-\alpha+\delta,1\}-\max\{\hb(X_{t})-\delta,0\}\\
 & =\min\{\hb(X_{t})+1-\alpha+\delta,1\}+\min\{\delta-\hb(X_{t}),0\}
\end{align*}
where (i) and (ii) follow by the fact that $U_{t}$ has the uniform
distribution on {[}0,1{]} and is independent of $X_{t}$ and $\hb(\cdot)$.
Thus, 
\begin{align}
 & P\left(|\tV_{t}^{*}|\leq\frac{1-\alpha}{2}+\delta\mid X_{t}\right)-(1-\alpha)\nonumber \\
 & =\min\{\hb(X_{t})+1-\alpha+\delta,1\}+\min\{\delta-\hb(X_{t}),0\}-(1-\alpha)\nonumber \\
 & =\min\{\hb(X_{t})+\delta,\alpha\}+\min\{\delta-\hb(X_{t}),0\}\nonumber \\
 & =\min\{\delta,\alpha-\hb(X_{t})\}+\hb(X_{t})+\min\{\delta,\hb(X_{t})\}-\hb(X_{t})\nonumber \\
 & =\min\{\delta,\alpha-\hb(X_{t})\}+\min\{\delta,\hb(X_{t})\}.\label{eq: eff gen part 1 eq 7}
\end{align}

We now consider the random mapping $\delta\mapsto g(\delta)=\min\{\delta,\alpha-\hb(X_{t})\}+\min\{\delta,\hb(X_{t})\}$,
where the randomness is from the randomness of $X_{t}$. Clearly,
\[
Eg(\delta)=\delta+\delta=2\delta\leq\delta\qquad{\rm for}\ \delta\in[-\alpha,0].
\]

For $\delta\in[0,\max\{\alpha-\hb(X_{t}),\hb(X_{t})\}]$, we have
$g(\delta)\geq\delta$. For $\delta\in[\max\{\alpha-\hb(X_{t}),\hb(X_{t})\},\alpha]$,
we have that 
\[
g(\delta)=\alpha\geq\delta.
\]

Hence, for any $\delta\in[0,\alpha]$, we have $g(\delta)\geq\delta$,
which implies $Eg(\delta)\geq\delta$. The proof is complete. 
\end{proof}
\begin{customlem}{S2}
\label{lem: eff gen part 2}Let Assumption \ref{assu: efficiency gen}
hold. Then $\sup_{(a,x)\in[0,1]\times\Xcal}|\hQ(a,x)-Q(a,x)|=o_{P}(1)$
and $\sup_{x\in\Xcal}|\hL(x)-L(x)|=o_{P}(1)$. 
\end{customlem}

\begin{proof}
Let $\varepsilon=\sup_{x\in\Xcal}\sup_{y\in\mathcal{Y}(x)}|\hF(y,x)-F(y,x)|$.
Fix an arbitrary $\delta>0$ and $x\in\Xcal$. Since $\hF(\cdot,x)$
is right-continuous and $\hQ(a,x)=\inf\{y:\ \hF(y,x)\geq a\}$ for
any $a\in[0,1]$, we have that $\hF(\hQ(a,x),x)=a$. For simplicity,
we write $\hF(y,x)$, $\hQ(a,x)$, $F(y,x)$ and $Q(a,x)$ as $\hF(y)$,
$\hQ(y)$, $F(y)$ and $Q(a)$, respectively whenever no confusion
arises.

We consider the event $\left\{ Q(a)>\hQ(a)+\delta\right\} $: 
\begin{multline*}
\left\{ Q(a)>\hQ(a)+\delta\right\} \subseteq\left\{ F(Q(a))\geq F(\hQ(a)+\delta)\right\} =\left\{ a\geq F(\hQ(a)+\delta)\right\} \\
\overset{\text{(i)}}{\subseteq}\left\{ a\geq F(\hQ(a))+C_{1}\delta\right\} \subseteq\left\{ a\geq\hF(\hQ(a))-\varepsilon+C_{1}\delta\right\} \overset{\text{(ii)}}{=}\left\{ \varepsilon\geq C_{1}\delta\right\} ,
\end{multline*}
where (i) follows by the fact that $F(b+\delta,x)=F(b,x)+\int_{b}^{b+\delta}f(z,x)dz\geq F(b,x)+\int_{b}^{b+\delta}C_{1}dz=F(b,x)+C_{1}\delta$
and (ii) follows by $\hF(\hQ(a))=a$. Similarly, we observe that 
\begin{multline*}
\left\{ \hQ(a)>Q(a)+\delta\right\} \subseteq\left\{ \hF(\hQ(a))\geq\hF(Q(a)+\delta)\right\} =\left\{ a\geq\hF(Q(a)+\delta)\right\} \\
\subseteq\left\{ a\geq F(Q(a)+\delta)-\varepsilon\right\} \subseteq\left\{ a\geq F(Q(a))+C_{1}\delta-\varepsilon\right\} =\left\{ \varepsilon\geq C_{1}\delta\right\} .
\end{multline*}

By the above two displays, we have that 
\[
\left\{ \left|\hQ(a)-Q(a)\right|>\delta\right\} =\left\{ Q(a)>\hQ(a)+\delta\right\} \bigcup\left\{ \hQ(a)>Q(a)+\delta\right\} \subseteq\left\{ \varepsilon\geq C_{1}\delta\right\} .
\]

Notice that the right-hand side $\left\{ \varepsilon\geq C_{1}\delta\right\} $
does not depend on $x$ or $a$. Therefore, 
\[
\left\{ \sup_{(a,x)\in[0,1]\times\Xcal}\left|\hQ(a,x)-Q(a,x)\right|>\delta\right\} \subseteq\left\{ \varepsilon\geq C_{1}\delta\right\} .
\]

Hence, 
\[
P\left(\sup_{(a,x)\in[0,1]\times\Xcal}\left|\hQ(a,x)-Q(a,x)\right|>\delta\right)\leq P\left(\varepsilon\geq C_{1}\delta\right)\overset{\text{(i)}}{=}o(1),
\]
where (i) follows by $\varepsilon=o_{P}(1)$. Since $\delta>0$ is
arbitrary, we have proved $\sup_{(a,x)\in[0,1]\times\Xcal}|\hQ(a,x)-Q(a,x)|=o_{P}(1)$. 

To show $\sup_{x\in\Xcal}|\hL(x)-L(x)|=o_{P}(1)$, we define $\eta=\sup_{(a,x)\in[0,1]\times\Xcal}|\hQ(a,x)-Q(a,x)|$.
We observe that 
\begin{align*}
\hL(x) & =\min_{z\in[0,\alpha]}\hQ(z+1-\alpha,x)-\hQ(z,x)\\
 & \leq\min_{z\in[0,\alpha]}\left(Q(z+1-\alpha,x)-Q(z,x)+2\eta\right)=L(x)+2\eta
\end{align*}
and 
\begin{align*}
\hL(x) & =\min_{z\in[0,\alpha]}\hQ(z+1-\alpha,x)-\hQ(z,x)\\
 & \geq\min_{z\in[0,\alpha]}\left(Q(z+1-\alpha,x)-Q(z,x)-2\eta\right)=L(x)-2\eta.
\end{align*}

Thus, $|\hL(x)-L(x)|\leq2\eta$. Since this holds for any $x\in\Xcal$,
we have $\sup_{x\in\Xcal}|\hL(x)-L(x)|\leq2\eta$. Because we have
proved $\eta=o_{P}(1)$, it follows that $\sup_{x\in\Xcal}|\hL(x)-L(x)|=o_{P}(1)$.
The proof is complete. 
\end{proof}
\begin{customlem}{S3}
\label{lem: eff gen part 3}Let Assumption \ref{assu: efficiency gen}
hold. Then $\hQ_{\Tcal_{2}}^{*}=(1-\alpha)/2+o_{P}(1)$. 
\end{customlem}

\begin{proof}
Fix an arbitrary $\delta\in(0,\alpha)$. Define the event 
\begin{multline*}
\Acal=\left\{ \max_{t\in\Tcal_{2}}|\hV_{t}^{*}-\tV_{t}^{*}|\leq\delta/2\right\} \\
\bigcap\left\{ \left||\Tcal_{2}|^{-1}\sum_{t\in\Tcal_{2}}\oneb\{|\tV_{t}^{*}|\leq(1-\alpha)/2+\delta/2\}-P\left(|\tV_{t}^{*}|\leq(1-\alpha)/2+\delta/2\right)\right|\leq\delta/4\right\} \\
\bigcap\left\{ \left||\Tcal_{2}|^{-1}\sum_{t\in\Tcal_{2}}\oneb\{|\tV_{t}^{*}|\leq(1-\alpha)/2-\delta/2\}-P\left(|\tV_{t}^{*}|\leq(1-\alpha)/2-\delta/2\right)\right|\leq\delta/4\right\} .
\end{multline*}

Since $\hQ_{\Tcal_{2}}^{*}$ is the $(1-\alpha)(1+|\Tcal_{2}|^{-1})$
sample quantile of $\{|\hV_{t}^{*}|\}_{t\in\Tcal_{2}}$, we have that
\begin{equation}
(1-\alpha)(1+|\Tcal_{2}|^{-1})-|\Tcal_{2}|^{-1}\leq|\Tcal_{2}|^{-1}\sum_{t\in\Tcal_{2}}\oneb\{|\hV_{t}^{*}|\leq\hQ_{\Tcal_{2}}^{*}\}\leq(1-\alpha)(1+|\Tcal_{2}|^{-1})+|\Tcal_{2}|^{-1}.\label{eq: lem eff gen part 3 eq 4}
\end{equation}

We consider the two events $\Mcal_{1}=\{\hQ_{\Tcal_{2}}^{*}>(1-\alpha)/2+\delta\}$
and $\Mcal_{2}=\{\hQ_{\Tcal_{2}}^{*}<(1-\alpha)/2-\delta\}$. We will
show three claims: $P(\Acal^{c})=o(1)$, $P(\Mcal_{1}\bigcap\Acal)=o(1)$
and $P(\Mcal_{2}\bigcap\Acal)=o(1)$. Then the desired result follows
by $P(|\hQ_{\Tcal_{2}}^{*}-(1-\alpha)/2|>\delta)=P(\Mcal_{1}\bigcup\Mcal_{2})$
together with
\[
P\left(\Mcal_{1}\bigcup\Mcal_{2}\right)\leq P(\Acal^{c})+P\left((\Mcal_{1}\bigcup\Mcal_{2})\bigcap\Acal\right)\leq P(\Acal^{c})+P(\Mcal_{1}\bigcap\Acal)+P(\Mcal_{2}\bigcap\Acal).
\]

We show these three claims in three steps. 

\textbf{Step 1:} show $P(\Acal^{c})=o(1)$.

Since $\{\tV_{t}^{*}\}_{t\in\Tcal_{2}}$ is independent, we have
\begin{multline*}
E\left[|\Tcal_{2}|^{-1}\sum_{t\in\Tcal_{2}}\oneb\{|\tV_{t}^{*}|\leq(1-\alpha)/2+\delta/2\}-P\left(|\tV_{t}^{*}|\leq(1-\alpha)/2+\delta/2\right)\right]^{2}\\
=|\Tcal_{2}|^{-2}\sum_{t\in\Tcal_{2}}E\left[\oneb\{|\tV_{t}^{*}|\leq(1-\alpha)/2+\delta/2\}-P\left(|\tV_{t}^{*}|\leq(1-\alpha)/2+\delta/2\right)\right]^{2}\overset{\text{(i)}}{\leq}\frac{1}{4|\Tcal_{2}|},
\end{multline*}
where (i) follows by the fact that $E(Z-P(Z=1))^{2}=P(Z=1)\cdot(1-P(Z=1))\leq\max_{x\in[0,1]}x(1-x)\leq1/4$
for any Bernoulli variable $Z$. Thus, 
\[
P\left(\left||\Tcal_{2}|^{-1}\sum_{t\in\Tcal_{2}}\oneb\{|\tV_{t}^{*}|\leq(1-\alpha)/2+\delta/2\}-P\left(|\tV_{t}^{*}|\leq(1-\alpha)/2+\delta/2\right)\right|>\delta/4\right)\leq\frac{1}{\delta|\Tcal_{2}|}=o(1).
\]

Similarly, we can show that 
\[
P\left(\left||\Tcal_{2}|^{-1}\sum_{t\in\Tcal_{2}}\oneb\{|\tV_{t}^{*}|\leq(1-\alpha)/2-\delta/2\}-P\left(|\tV_{t}^{*}|\leq(1-\alpha)/2-\delta/2\right)\right|>\delta/4\right)\leq\frac{1}{\delta|\Tcal_{2}|}=o(1).
\]

We notice that $|\hV_{t}^{*}-\tV_{t}^{*}|\leq|\hF(Y_{t},X_{t})-F(Y_{t},X_{t})|$
and thus
\[
P\left(\max_{t\in\Tcal_{2}}|\hV_{t}^{*}-\tV_{t}^{*}|>\delta/2\right)\leq P\left(\sup_{x\in\Xcal}\sup_{y\in\mathcal{Y}(x)}|\hF(y,x)-F(y,x)|>\delta/2\right)=o(1).
\]

Therefore, $P(\Acal^{c})=o(1)$. 

\textbf{Step 2:} show $P(\Mcal_{1}\bigcap\Acal)\rightarrow0$.

On the event $\Mcal_{1}\bigcap\Acal$, we have that 
\begin{align*}
 & |\Tcal_{2}|^{-1}\sum_{t\in\Tcal_{2}}\oneb\{|\tV_{t}^{*}|\leq(1-\alpha)/2+\delta/2\}\\
 & =|\Tcal_{2}|^{-1}\sum_{t\in\Tcal_{2}}\oneb\{|\hV_{t}^{*}|\leq(1-\alpha)/2+\delta/2+|\hV_{t}^{*}|-|\tV_{t}^{*}|\}\\
 & \overset{\text{(i)}}{\leq}|\Tcal_{2}|^{-1}\sum_{t\in\Tcal_{2}}\oneb\{|\hV_{t}^{*}|\leq(1-\alpha)/2+\delta/2+\delta/2\}\\
 & \overset{\text{(ii)}}{\leq}|\Tcal_{2}|^{-1}\sum_{t\in\Tcal_{2}}\oneb\{|\hV_{t}^{*}|\leq\hQ_{\Tcal_{2}}^{*}\}\overset{\text{(iii)}}{\leq}(1-\alpha)(1+|\Tcal_{2}|^{-1})+|\Tcal_{2}|^{-1},
\end{align*}
 where (i) follows by the definition of $\Acal$, (ii) follows by
the definition of $\Mcal_{1}$ and (iii) follows by \eqref{eq: lem eff gen part 3 eq 4}.
On the event $\Acal$, we have

\begin{footnotesize}
\[
|\Tcal_{2}|^{-1}\sum_{t\in\Tcal_{2}}\oneb\{|\tV_{t}^{*}|\leq(1-\alpha)/2+\delta/2\}\geq P\left(|\tV_{t}^{*}|\leq(1-\alpha)/2+\delta/2\right)-\delta/4\overset{\text{(i)}}{\geq}(1-\alpha)+\delta/2-\delta/4=(1-\alpha)+\delta/4,
\]
\end{footnotesize}
where (i) follows by Lemma \ref{lem: eff gen part 1}. 

The above two displays imply that on the event $\Mcal_{1}\bigcap\Acal$,
$(1-\alpha)+\delta/4\leq(1-\alpha)(1+|\Tcal_{2}|^{-1})+|\Tcal_{2}|^{-1}$,
which is $\delta\leq4(2-\alpha)/|\Tcal_{2}|$. Since $|\Tcal_{2}|\rightarrow\infty$
and $\delta>0$ is fixed, we have 
\[
P\left(\Mcal_{1}\bigcap\Acal\right)\leq\oneb\{\delta\leq4(2-\alpha)/|\Tcal_{2}|\}=o(1).
\]

\textbf{Step 3:} show $P(\Mcal_{2}\bigcap\Acal)\rightarrow0$.

The argument is similar to Step 2. On the event $\Mcal_{2}\bigcap\Acal$,
we have that 
\begin{align*}
 & |\Tcal_{2}|^{-1}\sum_{t\in\Tcal_{2}}\oneb\{|\tV_{t}^{*}|\leq(1-\alpha)/2-\delta/2\}\\
 & =|\Tcal_{2}|^{-1}\sum_{t\in\Tcal_{2}}\oneb\{|\hV_{t}^{*}|\leq(1-\alpha)/2-\delta/2+|\hV_{t}^{*}|-|\tV_{t}^{*}|\}\\
 & \geq|\Tcal_{2}|^{-1}\sum_{t\in\Tcal_{2}}\oneb\{|\hV_{t}^{*}|\leq(1-\alpha)/2-\delta/2-\delta/2\}\\
 & \geq|\Tcal_{2}|^{-1}\sum_{t\in\Tcal_{2}}\oneb\{|\hV_{t}^{*}|\leq\hQ_{\Tcal_{2}}^{*}\}\geq(1-\alpha)(1+|\Tcal_{2}|^{-1})-|\Tcal_{2}|^{-1}.
\end{align*}

On the event $\Acal$, we have
\begin{align*}
|\Tcal_{2}|^{-1}\sum_{t\in\Tcal_{2}}\oneb\{|\tV_{t}^{*}|\leq(1-\alpha)/2-\delta/2\} & \leq P\left(|\tV_{t}^{*}|\leq(1-\alpha)/2-\delta/2\right)+\delta/4\\
 & \overset{\text{(i)}}{\leq}(1-\alpha)-\delta/2+\delta/4=(1-\alpha)-\delta/4,
\end{align*}
where (i) follows by Lemma \ref{lem: eff gen part 1}. 

The above two displays imply that on the event $\Mcal_{2}\bigcap\Acal$,
$(1-\alpha)-\delta/4\geq(1-\alpha)(1+|\Tcal_{2}|^{-1})-|\Tcal_{2}|^{-1}$,
which is $\delta/4\leq\alpha/|\Tcal_{2}|$. Since $|\Tcal_{2}|\rightarrow\infty$
and $\delta>0$ is fixed, we have 
\[
P\left(\Mcal_{2}\bigcap\Acal\right)\leq\oneb\{\delta\leq4\alpha/|\Tcal_{2}|\}=o(1).
\]

The proof is complete.
\end{proof}

We are now ready to prove Theorem \ref{thm: effigency gen 2}. 

\begin{proof}[\textbf{Proof of Theorem \ref{thm: effigency gen 2}}]
Let $\varepsilon_{1}=\hQ_{\Tcal_{2}}^{*}-(1-\alpha)/2$, $\varepsilon_{2}=\sup_{y,x}|\hF(y,x)-F(y,x)|$,
$\varepsilon_{3}=\sup_{(a,x)\in[0,1]\times\Xcal}|\hQ(a,x)-Q(a,x)|$
and $\varepsilon_{4}=\sup_{x\in\Xcal}|\hL(x)-L(x)|$. For simplicity,
we write $\widehat{\Ccal_{(1-\alpha)}^{{\rm conf}}}$ instead of $\widehat{\Ccal_{(1-\alpha)}^{{\rm conf}}}(X_{T+1})$.
We proceed in two steps.

\textbf{Step 1:} show asymptotic conditional validity.

To show $P\left(Y_{T+1}\in\widehat{\Ccal_{(1-\alpha)}^{{\rm conf}}}\mid X_{T+1}\right)=1-\alpha+o_{P}(1)$,
it suffices to verify that $P(|\hV_{T+1}^{*}|\leq\hQ_{\Tcal_{2}}^{*}\mid X_{T+1})=1-\alpha+o_{P}(1)$.
We notice that 
\begin{align}
 & P\left(|\hV_{T+1}^{*}|\leq\hQ_{\Tcal_{2}}^{*}\mid X_{T+1}\right)\label{eq: thm 2nd eff eq 5}\\
 & =P\left(|\hF(Y_{T+1},X_{T+1})-\hb(X_{T+1})-(1-\alpha)/2|\leq\hQ_{\Tcal_{2}}^{*}\mid X_{T+1}\right)\nonumber \\
 & =P\left(\hb(X_{T+1})+(1-\alpha)/2-\hQ_{\Tcal_{2}}^{*}\leq\hF(Y_{T+1},X_{T+1})\leq\hb(X_{T+1})+(1-\alpha)/2+\hQ_{\Tcal_{2}}^{*}\mid X_{T+1}\right)\nonumber \\
 & =P\left(\hb(X_{T+1})-\varepsilon_{1}\leq\hF(Y_{T+1},X_{T+1})\leq\hb(X_{T+1})+(1-\alpha)+\varepsilon_{1}\mid X_{T+1}\right)\nonumber \\
 & =P\left(\hF(Y_{T+1},X_{T+1})\leq\hb(X_{T+1})+(1-\alpha)+\varepsilon_{1}\mid X_{T+1}\right)-P\left(\hF(Y_{T+1},X_{T+1})<\hb(X_{T+1})-\varepsilon_{1}\mid X_{T+1}\right).\nonumber 
\end{align}

Observe that 
\begin{multline*}
P\left(F(Y_{T+1},X_{T+1})<\hb(X_{T+1})-\varepsilon_{1}-\varepsilon_{2}\mid X_{T+1}\right)\leq P\left(\hF(Y_{T+1},X_{T+1})<\hb(X_{T+1})-\varepsilon_{1}\mid X_{T+1}\right)\\
\leq P\left(F(Y_{T+1},X_{T+1})<\hb(X_{T+1})-\varepsilon_{1}+\varepsilon_{2}\mid X_{T+1}\right).
\end{multline*}

Since $F(Y_{T+1},X_{T+1})$ is independent of $(\varepsilon_{1},\varepsilon_{2},\hb(X_{T+1}),X_{T+1})$
and has the uniform distribution on {[}0,1{]}, it follows that 
\[
\beta\left(\hb(X_{T+1})-\varepsilon_{1}-\varepsilon_{2}\right)\leq P\left(\hF(Y_{T+1},X_{T+1})<\hb(X_{T+1})-\varepsilon_{1}\mid X_{T+1}\right)\leq\beta\left(\hb(X_{T+1})-\varepsilon_{1}+\varepsilon_{2}\right),
\]
where 
\[
\beta(z)=\begin{cases}
1 & {\rm if}\ z>1\\
0 & {\rm if}\ z<0\\
z & {\rm otherwise}.
\end{cases}
\]

Clearly, $|\beta(z_{1})-\beta(z_{2})|\leq|z_{1}-z_{2}|$ for any $z_{1},z_{2}\in\RR$.
Thus, $|\beta(\hb(X_{T+1})-\varepsilon_{1}+\varepsilon_{2})-\beta(\hb(X_{T+1}))|\leq|-\varepsilon_{1}+\varepsilon_{2}|$
and $|\beta(\hb(X_{T+1})-\varepsilon_{1}-\varepsilon_{2})-\beta(\hb(X_{T+1}))|\leq|-\varepsilon_{1}-\varepsilon_{2}|$.
This means that 
\[
\left|P\left(\hF(Y_{T+1},X_{T+1})<\hb(X_{T+1})-\varepsilon_{1}\mid X_{T+1}\right)-\beta\left(\hb(X_{T+1})\right)\right|\leq|\varepsilon_{1}|+\varepsilon_{2}.
\]

Similarly, 
\[
\left|P\left(\hF(Y_{T+1},X_{T+1})\leq\hb(X_{T+1})+1-\alpha+\varepsilon_{1}\mid X_{T+1}\right)-\beta\left(\hb(X_{T+1})+1-\alpha\right) \right|\leq|\varepsilon_{1}|+\varepsilon_{2}.
\]

By $\hb(X_{T+1})\in[0,\alpha]$, we have $\beta(\hb(X_{T+1}))=\hb(X_{T+1})$ and $\beta(\hb(X_{T+1})+1-\alpha)=\hb(X_{T+1})+1-\alpha$.
Hence, the above two displays imply that 
\begin{footnotesize}
\begin{multline*}
\left|P\left(\hF(Y_{T+1},X_{T+1})\leq\hb(X_{T+1})+(1-\alpha)+\varepsilon_{1}\mid X_{T+1}\right)-P\left(\hF(Y_{T+1},X_{T+1})<\hb(X_{T+1})-\varepsilon_{1}\mid X_{T+1}\right)-(1-\alpha)\right|\\
\leq2|\varepsilon_{1}|+2\varepsilon_{2}.
\end{multline*}
\end{footnotesize}
By \eqref{eq: thm 2nd eff eq 5}, we have 
\[
\left|P\left(|\hV_{T+1}^{*}|\leq\hQ_{\Tcal_{2}}^{*}\mid X_{T+1}\right)-(1-\alpha)\right|\leq2|\varepsilon_{1}|+2\varepsilon_{2}.
\]

Since $\varepsilon_{1}$ and $\varepsilon_{2}$ are $o_{P}(1)$, we
have $\left|P\left(|\hV_{T+1}^{*}|\leq\hQ_{\Tcal_{2}}^{*}\mid X_{T+1}\right)-(1-\alpha)\right|=o_{P}(1)$. 

\textbf{Step 2:} show asymptotic efficiency.

We can rewrite the interval $\widehat{\Ccal_{(1-\alpha)}^{{\rm conf}}}=\{y:\ |\hF(y,X_{T+1})-\hb(X_{T+1})-(1-\alpha)/2|\leq\hQ_{\Tcal_{2}}^{*}\}$
as 
\[
\widehat{\Ccal_{(1-\alpha)}^{{\rm conf}}}=\left\{ y:\ \hb(X_{T+1})+(1-\alpha)/2-\hQ_{\Tcal_{2}}^{*}\leq\hF(y,X_{T+1})\leq\hb(X_{T+1})+(1-\alpha)/2+\hQ_{\Tcal_{2}}^{*}\right\} .
\]

In other words, we can write it as 
\[
\widehat{\Ccal_{(1-\alpha)}^{{\rm conf}}}=\left[\hQ\left(\beta(\hb(X_{T+1})+(1-\alpha)/2-\hQ_{\Tcal_{2}}^{*}),X_{T+1}\right),\hQ\left(\beta(\hb(X_{T+1})+(1-\alpha)/2+\hQ_{\Tcal_{2}}^{*}),X_{T+1}\right)\right].
\]

We can now compute the length of $\widehat{\Ccal_{(1-\alpha)}^{{\rm conf}}}$.
We observe
\begin{align}
\mu\left(\widehat{\Ccal_{(1-\alpha)}^{{\rm conf}}}\right) & =\hQ\left(\beta(\hb(X_{T+1})+(1-\alpha)/2+\hQ_{\Tcal_{2}}^{*}),X_{T+1}\right)-\hQ\left(\beta(\hb(X_{T+1})+(1-\alpha)/2-\hQ_{\Tcal_{2}}^{*}),X_{T+1}\right)\nonumber \\
 & \overset{\text{(i)}}{=}\hQ\left(\beta(\hb(X_{T+1})+1-\alpha+\varepsilon_{1}),X_{T+1}\right)-\hQ\left(\beta(\hb(X_{T+1})-\varepsilon_{1}),X_{T+1}\right)\nonumber \\
 & \leq Q\left(\beta(\hb(X_{T+1})+1-\alpha+\varepsilon_{1}),X_{T+1}\right)-Q\left(\beta(\hb(X_{T+1})-\varepsilon_{1}),X_{T+1}\right)+2\varepsilon_{3},\label{eq: thm 2nd eff eq 9}
\end{align}
where (i) follows by $\hQ_{\Tcal_{2}}^{*}=(1-\alpha)/2+\varepsilon_{1}$.

We notice that for any $a_{1},a_{2}\in[0,1]$ with $a_{1}>a_{2}$
and for any $x\in\Xcal$,
\[
Q(a_{1},x)-Q(a_{2},x)=\int_{a_{2}}^{a_{1}}\left(\frac{\partial Q(z,x)}{\partial z}\right)dz=\int_{a_{2}}^{a_{1}}\left(\frac{1}{f(Q(z,x),x)}\right)dz\leq\int_{a_{2}}^{a_{1}}\left(\frac{1}{C_{1}}\right)dz=(a_{1}-a_{2})/C_{1}.
\]
 
Therefore, 
\[
\sup_{a_{1},a_{2}\in[0,1],\ a_{1}\neq a_{2}}\sup_{x\in\Xcal}\left|\frac{Q(a_{1},x)-Q(a_{2},x)}{a_{1}-a_{2}}\right|\leq1/C_{1}.
\]

Since $|\beta(z_{1})-\beta(z_{2})|\leq|z_{1}-z_{2}|$ for any $z_{1},z_{2}\in\RR$,
it follows that 
\begin{multline*}
\left|Q\left(\beta(\hb(X_{T+1})+1-\alpha+\varepsilon_{1}),X_{T+1}\right)-Q\left(\beta(\hb(X_{T+1})+1-\alpha),X_{T+1}\right)\right|\\
\leq\left|\beta(\hb(X_{T+1})+1-\alpha+\varepsilon_{1})-\beta(\hb(X_{T+1})+1-\alpha)\right|/C_{1}\leq|\varepsilon_{1}|/C_{1}
\end{multline*}
 and 
\[
\left|Q\left(\beta(\hb(X_{T+1})-\varepsilon_{1}),X_{T+1}\right)-Q\left(\beta(\hb(X_{T+1})),X_{T+1}\right)\right|\leq|\varepsilon_{1}|/C_{1}.
\]

The above two displays and \eqref{eq: thm 2nd eff eq 9} imply 
\begin{align*}
\mu\left(\widehat{\Ccal_{(1-\alpha)}^{{\rm conf}}}\right) & \leq2|\varepsilon_{1}|/C_{1}+2\varepsilon_{3}+Q\left(\beta(\hb(X_{T+1})+1-\alpha),X_{T+1}\right)-Q\left(\beta(\hb(X_{T+1})),X_{T+1}\right)\\
 & \overset{\text{(i)}}{=}2|\varepsilon_{1}|/C_{1}+2\varepsilon_{3}+Q\left(\hb(X_{T+1})+1-\alpha,X_{T+1}\right)-Q\left(\hb(X_{T+1}),X_{T+1}\right)\\
 & \leq2|\varepsilon_{1}|/C_{1}+4\varepsilon_{3}+\hQ\left(\hb(X_{T+1})+1-\alpha,X_{T+1}\right)-\hQ\left(\hb(X_{T+1}),X_{T+1}\right)\\
 & =2|\varepsilon_{1}|/C_{1}+4\varepsilon_{3}+\hL(X_{T+1})\\
 & \leq2|\varepsilon_{1}|/C_{1}+4\varepsilon_{3}+\varepsilon_{4}+L(X_{T+1})\overset{\text{(ii)}}{=}L(X_{T+1})+o_{P}(1),
\end{align*}
where (i) follows by $\hb(X_{T+1})\in[0,\alpha]$ and (ii) follows
by $\varepsilon_{3}=o_{P}(1)$ and $\varepsilon_{4}=o_{P}(1)$ (Lemma
\ref{lem: eff gen part 2}) as well as $\varepsilon_{1}=o_{P}(1)$
(Lemma \ref{lem: eff gen part 3}). The desired result follows by
\begin{align*}
\mu\left(\Ccal_{(1-\alpha)}^{{\rm opt}}(X_{T+1})\right) & =\min_{F(z_{1},X_{T+1})-F(z_{2},X_{T+1})\geq1-\alpha}\ z_{1}-z_{2}\\
 & =\min_{F(z_{1},X_{T+1})-F(z_{2},X_{T+1})=1-\alpha}\ z_{1}-z_{2}\\
 & =\min_{z\in[0,\alpha]}Q(z+1-\alpha,X_{T+1})-Q(z,X_{T+1})\\
 & =L(X_{T+1}).
\end{align*}
\end{proof}

\subsection{Proof of Theorem \ref{thm: efficiency}}

For simplicity, we may omit $X_{T+1}$ and $\alpha$ when no confusion
can arise. For example, we write $F(y)$, $Q(y)$, $f(y)$ and $b$
rather than $F(y,X_{T+1})$, $Q(y,X_{T+1})$, $f(y,X_{T+1})$ and
$b(X_{T+1},\alpha)$, respectively. 

By Lemma \ref{lem: opt score}, we have that 
\[
\Ccal_{(1-\alpha)}^{{\rm opt}}(X_{T+1})=\Ccal_{(1-\alpha)}^{{\rm conf}}(X_{T+1})=\left\{ y:\ \left|F(y)-b-\frac{1-\alpha}{2}\right|\leq Q_{\psi}(1-\alpha)\right\} ,
\]
where $Q_{\psi}(1-\alpha)$ is the $(1-\alpha)$ quantile of $V_{t}^{*}=F(Y_{t},X_{t})-b(X_{t},\alpha)-\frac{1-\alpha}{2}$.
Again by Lemma \ref{lem: opt score}, $Q_{\psi}(1-\alpha)=\frac{1-\alpha}{2}$.
Therefore, 
\begin{equation}
\Ccal_{(1-\alpha)}^{{\rm opt}}(X_{T+1})=\left\{ y:\ b\leq F(y)\leq b+1-\alpha\right\} .\label{eq: thm opt eq 6}
\end{equation}

On the other hand, 
\begin{align*}
\widehat{\Ccal_{(1-\alpha)}^{{\rm conf}}}(X_{T+1}) & =\left\{ y:\ \hb(X_{T+1},\alpha)+\frac{1-\alpha}{2}-\hQ_{\mathcal{T}_{2}}^{*}\leq\hF(y,X_{T+1})\leq\hb(X_{T+1},\alpha)+\frac{1-\alpha}{2}+\hQ_{\mathcal{T}_{2}}^{*}\right\} \\
 & =\left\{ y:\ b+\varepsilon_{1}(y)\leq F(y,X_{T+1})\leq b+1-\alpha+\varepsilon_{2}(y)\right\} ,
\end{align*}
where $\varepsilon_{1}(y)=\hb(X_{T+1},\alpha)-b+\frac{1-\alpha}{2}-\hQ_{\mathcal{T}_{2}}^{*}+F(y,X_{T+1})-\hF(y,X_{T+1})$
and $\varepsilon_{2}(y)=\varepsilon_{1}(y)+2\hQ_{\mathcal{T}_{2}}^{*}-(1-\alpha)$.

The rest of the proof proceeds in two steps. 

\textbf{Step 1:} show that $\hQ_{\mathcal{T}_{2}}^{*}=(1-\alpha)/2+o_{P}(1)$.

Notice that 
\[
\left| |\hV_{t}^{*}|-|V_{t}^{*}| \right|\leq\left|\hF(Y_{t},X_{t})-F(Y_{t},X_{t})\right|+\left|\hb(X_{t},\alpha)-b(X_{t},\alpha)\right|.
\]

By the elementary inequality $(a+b)^{2}\leq2a^{2}+2b^{2}$, we have
that 

\begin{footnotesize}
\[
|\mathcal{T}_{2}|^{-1}\sum_{t\in\mathcal{T}_{2}}(|\hV_{t}^{*}|-|V_{t}^{*}|)^{2}\leq2|\mathcal{T}_{2}|^{-1}\sum_{t\in\mathcal{T}_{2}}\left(\hF(Y_{t},X_{t})-F(Y_{t},X_{t})\right)^{2}+2|\mathcal{T}_{2}|^{-1}\sum_{t\in\mathcal{T}_{2}}\left(\hb(X_{t},\alpha)-b(X_{t},\alpha)\right)^{2}=o_{P}(1).
\]
\end{footnotesize}

We now show that $G_{*}(\cdot)$ is Lipschitz. Fix any $y_{1},y_{2}$
in the support of $V_{t}^{*}$ such that $y_{1}<y_{2}$. Notice that
\begin{align*}
 & P(y_{1}\leq |V_{t}^{*}|\leq y_{2}\mid X_{t})\\
 & =P\left(y_{1}\leq\left|U_{t}-b(X_{t},\alpha)-\frac{1}{2}(1-\alpha)\right|\leq y_{2}\mid X_{t}\right)\\
 & =P\left(y_{1}\leq U_{t}-b(X_{t},\alpha)-\frac{1}{2}(1-\alpha)\leq y_{2}\mid X_{t}\right)+P\left(y_{1}\leq-\left[U_{t}-b(X_{t},\alpha)-\frac{1}{2}(1-\alpha)\right]\leq y_{2}\mid X_{t}\right)\\
 & \overset{\text{(i)}}{\leq}(y_{2}-y_{1})+(y_{2}-y_{1})\leq2(y_{2}-y_{1}),
\end{align*}
where (i) follows by the fact that conditional on $X_{t}$, $U_{t}$
follows the uniform distribution on $(0,1)$. Thus, 
\[
G_{*}(y_{2})-G_{*}(y_{1})=P(y_{1}\leq |V_{t}^{*}| \leq y_{2})\leq2(y_{2}-y_{1}).
\]

Therefore, $\sup_{y_{1}\neq y_{2}}|G_{*}(y_{2})-G_{*}(y_{1})|/|y_{2}-y_{1}|\leq2$.
By the same argument as the in the proof of Lemma \ref{lem: basic}, 
\[
\sup_{v\in\RR}\left|\tG_{*}(v)-G_{*}(v)\right|=o_{P}(1).
\]

By the continuity of $G_{*}(\cdot)$, we have that $\hQ_{\mathcal{T}_{2}}^{*}=G_{*}^{-1}(1-\alpha)+o_{P}(1)$
(since $\hQ_{\mathcal{T}_{2}}^{*}$ is the $(1-\alpha)(1+1/|\mathcal{T}_{2}|)$
quantile of $\tG_{*}(\cdot)$). Notice that $G_{*}^{-1}(1-\alpha)=Q_{\psi}(1-\alpha)$.
By Lemma \ref{lem: opt score}, $Q_{\psi}(1-\alpha)=(1-\alpha)/2$.
This means that $\hQ_{\mathcal{T}_{2}}^{*}=(1-\alpha)/2+o_{P}(1)$. 

\textbf{Step 2:} derive the final result.

By Step 1 and the assumptions that $\hb(X_{T+1},\alpha)-b=o_{P}(1)$
and $\sup_{y\in\RR}\left|\hF(y,X_{T+1})-F(y,X_{T+1})\right|=o_{P}(1)$,
we have that $\bar{\varepsilon}_{1}:=\sup_{y\in\RR}|\varepsilon_{1}(y)|=o_{P}(1)$
and $\bar{\varepsilon}_{2}:=\sup_{y\in\RR}|\varepsilon_{2}(y)|=o_{P}(1)$.
Define $H_{1}=\{y:\ b-\bar{\varepsilon}_{1}\leq F(y)\leq b+1-\alpha+\bar{\varepsilon}_{2}\}$
and $H_{2}=\{y:\ b+\bar{\varepsilon}_{1}\leq F(y)\leq b+1-\alpha-\bar{\varepsilon}_{2}\}$.
Clearly, 
\begin{equation}
H_{2}\subseteq\widehat{\Ccal_{(1-\alpha)}^{{\rm conf}}}(X_{T+1})\subseteq H_{1}\qquad\text{almost surely.}\label{eq: thm opt eq 7}
\end{equation}

On the other hand, we observe that $H_{1}$ is an interval that can
be written as 
\[
H_{1}=\left[Q\left(\max\{b-\bar{\varepsilon}_{1},0\}\right),\ Q\left(\min\{b+1-\alpha+\bar{\varepsilon}_{2},1\}\right)\right].
\]

Since $P(|Y_{T+1}|\leq C_{2}\mid X_{T+1})=1$ and $F(\cdot)$ is strictly
increasing, $Q(0)$ and $Q(1)$ are well defined and satisfy $\max\{|Q(0)|,|Q(1)|\}\leq C_{2}$
almost surely. 

By \eqref{eq: thm opt eq 6}, we can write $\Ccal_{(1-\alpha)}^{{\rm opt}}(X_{T+1})$
as an interval
\[
\Ccal_{(1-\alpha)}^{{\rm opt}}(X_{T+1})=[Q(b),Q(b+1-\alpha)].
\]

Therefore, 
\[
\mu\left(H_{1}\triangle\Ccal_{(1-\alpha)}^{{\rm opt}}(X_{T+1})\right)\leq\left|Q\left(\max\{b-\bar{\varepsilon}_{1},0\}\right)-Q(b)\right|+\left|Q\left(\min\{b+1-\alpha+\bar{\varepsilon}_{2},1\}\right)-Q(b+1-\alpha)\right|.
\]

Notice that $dQ(u)/du=1/f(Q(u))$. By assumption, the density is bounded
below by $C_{1}$ on the support of $Y_{T+1}\mid X_{T+1}$. It follows
that $|dQ(u)/du|$ is uniformly bounded by $1/C_{1}$. Thus, 
\[
\left|Q\left(\max\{b-\bar{\varepsilon}_{1},0\}\right)-Q(b)\right|\leq\frac{1}{C_{1}}\cdot|\max\{b-\bar{\varepsilon}_{1},0\}-b|\leq\frac{1}{C_{1}}\cdot\bar{\varepsilon}_{1}
\]
and similary
\[
\left|Q\left(\min\{b+1-\alpha+\bar{\varepsilon}_{2},1\}\right)-Q(b+1-\alpha)\right|\leq\frac{1}{C_{1}}\cdot\bar{\varepsilon}_{2}.
\]

The above three displays imply 
\[
\mu\left(H_{1}\triangle\Ccal_{(1-\alpha)}^{{\rm opt}}(X_{T+1})\right)\leq(\bar{\varepsilon}_{1}+\bar{\varepsilon}_{2})C_{1}^{-1}.
\]

Similarly, we can show that 
\[
\mu\left(H_{2}\triangle\Ccal_{(1-\alpha)}^{{\rm opt}}(X_{T+1})\right)\leq(\bar{\varepsilon}_{1}+\bar{\varepsilon}_{2})C_{1}^{-1}.
\]

By \eqref{eq: thm opt eq 7}, we have that, almost surely
\[
\widehat{\Ccal_{(1-\alpha)}^{{\rm conf}}}(X_{T+1})\triangle\Ccal_{(1-\alpha)}^{{\rm opt}}(X_{T+1})\subseteq\left(H_{1}\triangle\Ccal_{(1-\alpha)}^{{\rm opt}}(X_{T+1})\right)\bigcup\left(H_{2}\triangle\Ccal_{(1-\alpha)}^{{\rm opt}}(X_{T+1})\right).
\]

The above three displays imply 
\[
\mu\left(\widehat{\Ccal_{(1-\alpha)}^{{\rm conf}}}(X_{T+1})\triangle\Ccal_{(1-\alpha)}^{{\rm opt}}(X_{T+1})\right)\leq2(\bar{\varepsilon}_{1}+\bar{\varepsilon}_{2})C_{1}^{-1}.
\]

The desired result follows by $\bar{\varepsilon}_{1}=o_{P}(1)$ and
$\bar{\varepsilon}_{2}=o_{P}(1)$.

\section{Time series discussion}
\label{app: time series discussion}
For time series data $\{Z_{t}\}_{t=1}^{T+1}$ with $Z_{t}=(X_{t},Y_{t})$,
it is often plausible that these $T+1$ observations are not independent.
Here, we assume that data is strictly stationary, i.e., for any $m>1$,
the distribution $(Z_{t-m},Z_{t-m+1},\dots,Z_{t-1})$ does not depend
on $t$. This is a common assumption in the time series literature.
Although the data is not independent, it is often not strongly dependent
either. Usually, we work with various notions of weak dependence.
A popular way of defining weak dependence is in terms of mixing conditions.
There are numerous mixing conditions, see, for example,
\cite{bradley2005basic,bradley2007introduction,dedecker2007weak}.
We focus on the $\beta$-mixing condition (also known as the absolute
regularity condition): for any $m>1$,
\[
\beta(m)=\frac{1}{2} \|P_{\{Z_{t}\}_{t\leq s},\ \{Z_{t}\}_{t\geq s+m}}-P_{\{Z_{t}\}_{t\leq s}}\otimes P_{\{Z_{t}\}_{t\geq s+m}}\|_{TV},
\]
where $P_{\{Z_{t}\}_{t\leq s}}$ denotes the probability measure of
$\{Z_{t}\}_{t\leq s}$, $P_{\{Z_{t}\}_{t\geq s+m}}$ denotes the probability
measure of $\{Z_{t}\}_{t\geq s+m}$ and $P_{\{Z_{t}\}_{t\leq s},\ \{Z_{t}\}_{t\geq s+m}}$
denotes the probability measure of the joint random components $(\{Z_{t}\}_{t\leq s},\ \{Z_{t}\}_{t\geq s+m})$.
Here, $\otimes$ denotes the product measure and $\|\cdot\|_{TV}$
is the total-variation norm. Since the data is strictly stationary, the
above definition does not depend on $s$. We borrow the above definition
of Section 1.6 of \cite{rio2017asymptotic}, but equivalent definitions
can be found in \cite{bradley2005basic,bradley2007introduction}
among others.\footnote{To see that these definitions are equivalent, one can
find details in Theorem 3.29 of \cite{bradley2007introduction}.}

We say that the sequence $\{Z_{t}\}$ is $\beta$-mixing if $\beta(m)\rightarrow0$
as $m\rightarrow\infty$. The $\beta$-mixing condition captures the idea that observations
that are far apart in time become nearly independent. As $m$ increases,
$\{Z_{t}\}_{t\leq s}$ and $\{Z_{t}\}_{t\geq s+m}$ become more independent,
in the sense that the joint distribution $P_{\{Z_{t}\}_{t\leq s},\ \{Z_{t}\}_{t\geq s+m}}$
is close to the product measure of the marginal distributions $P_{\{Z_{t}\}_{t\leq s}}\otimes P_{\{Z_{t}\}_{t\geq s+m}}$.

The $\beta$-mixing condition is satisfied for a large class of stochastic
processes. The simplest examples are perhaps $m$-dependent processes,
which satisfy that $\{Z_{j}\}_{j\leq t}$ and $\{Z_{j}\}_{j\geq s}$
are independent as long as $s-t\geq m$ for some fixed $m$. Moving
average processes are $m$-dependent. Autoregressive moving average
(ARMA) processes with independent errors are $\beta$-mixing. In general,
strictly stationary Markov chains that are Harris recurrent and aperiodic
are $\beta$-mixing \citep[e.g.,][]{bradley2005basic,meyn2012markov}.
Several stochastic volatility models for asset returns, including the
popular generalized autoregressive conditionally heteroskedastic (GARCH)
models are also $\beta$-mixing with $\beta(m)$ decaying exponentially
with $m$ \citep[e.g.,][]{boussama1998ergodicite,carrasco2002mixing,francq2006mixing}.

Now we consider the problem of empirical risk minimization mentioned
in Section \ref{sec:theory}. Let $\mathcal{F}$ be a model, i.e., a class of functions
of $Z_{t}=(X_{t},Y_{t})$. Define $F^{*}=\arg\min_{f\in\mathcal{F}}R_{T+1}(f)$,
where $R_{T+1}(f)=(T+1)^{-1}\sum_{t=1}^{T+1}E[L(Z_{t},f)]$, where
$L$ is a loss function. Let $\hF=\arg\min_{f\in\mathcal{F}}\hat{R}_{T+1}(f)$,
where $\hat{R}_{T+1}(f)=(T+1)^{-1}\sum_{t=1}^{T+1}L(Z_{t},f)$. Suppose
that the following entropy condition with brackets holds: 
\[
\int_{0}^{1}\sqrt{\varepsilon^{-1}\log N_{[]}(\varepsilon,L(\mathcal{F}),\|\cdot\|_{1,P})}d\varepsilon<\infty,
\]
where $N_{[]}$ is the bracketing number (see \cite{van1996weak}),
$L(\mathcal{F})$ is the class $\{L(Z_{t},f):\ f\in\mathcal{F}\}$
and $\|\cdot\|_{1,P}$ is the $L_{1}$-norm $\|f\|_{1,P}=E|f(Z_{t})|$.
By Theorem 8.3 of \cite{rio2017asymptotic}, $\sup_{f\in\mathcal{F}}|\hat{R}_{T+1}(f)-R_{T+1}(f)|=O_{P}(T^{-1/2})$
as long as $\sum_{m=1}^{\infty}\beta(m)<\infty$. (Similar results
for empirical processes of dependent data can be found in \cite{dedecker2002maximal}.)
By the usual arguments, it follows that $0\leq R_{T+1}(\hF)-R_{T+1}(F^{*})\leq o_{P}(1)$. Suppose that the
risk function is convex in a neighborhood of $F^{*}$: there exist
$C_{1},C_{2}>0$ such that $R_{T+1}(f)-R_{T+1}(F^{*})\geq C_{2}\|f-F^{*}\|_{\sup}^{2}$
whenever $\|f-F^{*}\|_{\sup}\leq C_{1}$ and $f\in\mathcal{F}$, where
$\|f\|_{\sup}=\sup_{z}|f(z)|$. Then $\sup_{y,x}|\hF(y,x)-F^{*}(y,x)|=\sup_{z}|\hF(z)-F^{*}(z)|=o_{P}(1)$.
This implies the consistency requirement in Assumption \ref{assu: basic}. Importantly,
$F^{*}$ does not need to be the true CDF $F$ because $F$ may or may not be in $\mathcal{F}$.

For the popular linear QR model, we establish a more concrete result; similar results can be established for DR.
Suppose that $X_{t}\in\RR^{d}$ for a fixed $d$. Let $\hat{\gamma}(u)=\arg\min_{\gamma\in\Gamma}\sum_{t=1}^{T+1}\rho_{u}(Y_{t}-X_{t}^{\top}\gamma)$
for $u\in[c_{T},1-c_{T}]$, where $\Gamma\subset\RR^{d}$ is a compact
set and $c_{T}>0$ is either a small constant or a sequence tending
to zero. Define
\[
\hF(y,x)=c_{T}+\int_{c_{T}}^{1-c_{T}}\oneb\{x^{\top}\hat{\gamma}(u)\leq y\}du.
\]

Let $\|\cdot\|_{2}$ denote the Euclidean norm in $\RR^{d}$. We have
the following result.

\begin{customthm}{S1}
\label{thm: linear QR consistency} Assume that the data $(X_{t},Y_{t})$ is strictly
stationary. Let $\gamma^{*}(u)=\arg\min_{\gamma\in\Gamma}E\rho_{u}(Y_{t}-X_{t}^{\top}\gamma)$
with $\rho_{u}(a)=a(u-\oneb\{a\leq0\})$. Define 
\[
F^{*}(y,x)=c_{T}+\int_{c_{T}}^{1-c_{T}}\oneb\{x^{\top}\gamma^{*}(u)\leq y\}du.
\]
Suppose that the following conditions hold:
\begin{enumerate}
\item There exists a constant $C_{1}>0$ such that $\|X_{t}\|_{2}\leq C_{1}$
and $|Y_{t}|\leq C_{1}$ almost surely. 
\item The $\beta$-mixing coefficient of $(X_{t},Y_{t})$ satisfies $\sum_{m=1}^{\infty}\beta(m)<\infty$. 
\item There exists a function $h(\cdot)$ such that $\lim_{\delta\rightarrow0}h(\delta)=0$
and $|F^{*}(y_{1},x)-F^{*}(y_{2},x)|\leq h(|y_{1}-y_{2}|)$ for any
$(y_{1},y_{2})$ and any $x$ with $\|x\|_{2}\leq C_{1}$.
\item $f(y,x)=\partial F(y,x)/\partial y$ exists and there exists a constant
$C_{2}>0$ such that $f(y,x)\geq C_{2}$ for any $x$ and any $y\in[s_{1}(x),\ s_{2}(x)]$,
where $[s_{1}(x),\ s_{2}(x)]$ is the support of the distribution
$Y_{t}\mid X_{t}=x$.
\item the smallest eigenvalue of $E(X_{t}X_{t}^{\top})$ is bounded below
by a constant $C_{3}>0$.
\end{enumerate}
Then $\sup_{y\in\RR,\ \|x\|_{2}\leq C_{1}}|\hF(y,x)-F^{*}(y,x)|=o_{P}(1)$.
\end{customthm}

Theorem \ref{thm: linear QR consistency} establishes the uniform
consistency of $\hF$, which guarantees the consistency requirement
in Assumption \ref{assu: basic} . Notice that Theorem \ref{thm: linear QR consistency}
does not assume that $F^{*}$ is the true conditional distribution
function $F$. It thus generalizes the analysis of QR under misspecification in \cite{angrist2006quantile} to time series settings.

The assumptions of Theorem \ref{thm: linear QR consistency}
are relatively mild. The boundedness of $X_{t}$ and $Y_{t}$ can
be relaxed with extra technical arguments. The summability of $\beta$-mixing
coefficients holds if $\beta(m)$ decays exponentially. The third
assumption says that $F^{*}(y,x)$ is uniformly continuous in $y$.
The last assumption states that the true conditional density of $Y_{t}\mid X_{t}$
is bounded away from zero on the support.

\begin{proof}[\textbf{Proof of Theorem \ref{thm: linear QR consistency}}]
We proceed in two steps.

\textbf{Step 1:} show that $\sup_{u\in[c_{T},1-c_{T}]}\|\hat{\gamma}(u)-\gamma^{*}(u)\|_{2}=o_{P}(1)$.

Let $R(\gamma,u)=E\rho_{u}(Y_{t}-X_{t}^{\top}\gamma)$ and $\hat{R}(\gamma,u)=(T+1)^{-1}\sum_{t=1}^{T+1}\rho_{u}(Y_{t}-X_{t}^{\top}\gamma)$.
For any $u_{1},u_{2}\in[c_{T},1-c_{T}]$ and $\gamma_{1},\gamma_{2}\in\Gamma$,
we observe 
\begin{align*}
 & \left|\rho_{u_{1}}(y-x^{\top}\gamma_{1})-\rho_{u_{2}}(y-x^{\top}\gamma_{2})\right|\\
 & \leq\left|\rho_{u_{1}}(y-x^{\top}\gamma_{1})-\rho_{u_{1}}(y-x^{\top}\gamma_{2})\right|+\left|\rho_{u_{1}}(y-x^{\top}\gamma_{2})-\rho_{u_{2}}(y-x^{\top}\gamma_{2})\right|\\
 & \leq\max\{u_{1},1-u_{1}\}\cdot|x^{\top}\gamma_{1}-x^{\top}\gamma_{2}|+|u_{1}-u_{2}|\cdot|y-x^{\top}\gamma_{2}|\\
 & \overset{\text{(i)}}{\leq}C_{1}\|\gamma_{1}-\gamma_{2}\|_{2}+\left(1+\sup_{\gamma\in\Gamma}\|\gamma\|_{2}\right)C_{1}|u_{1}-u_{2}|,
\end{align*}
where (i) follows by $|y-x^{\top}\gamma_{2}|\leq C_{1}+C_{1}\sup_{\gamma\in\Gamma}\|\gamma\|_{2}$.
Since $\Gamma$ is compact and $d$ is fixed, $\sup_{\gamma\in\Gamma}\|\gamma\|_{2}$
is bounded by a positive constant. Hence, $(\gamma,u)\mapsto\rho_{u}(y-x^{\top}\gamma)$
is Lipschitz. By Theorem 2.7.11 of \cite{van1996weak} and the usual
covering number bounds for Euclidean balls (e.g., Corollary 4.2.13
of \cite{vershynin2018high}), we have that for any norm $\|\cdot\|$,
the bracketing number satisfies
\[
N_{[]}(\varepsilon,\mathcal{G},\|\cdot\|)\leq(K_{1}/\varepsilon)^{d},
\]
where $K_{1}\geq1$ is a constant depending only on $C_{1}$, $\sup_{\gamma\in\Gamma}\|\gamma\|_{2}$
and $d$ and $\mathcal{G}$ is the class of functions $\rho_{u}(y-x^{\top}\gamma)$
with $(\gamma,u)\in\Gamma\times[c_{1},1-c_{2}]$. Notice that 
\begin{align*}
\int_{0}^{1}\sqrt{\varepsilon^{-1}\log N_{[]}(\varepsilon,\mathcal{G},\|\cdot\|_{1,P})}d\varepsilon & \leq\int_{0}^{1}\sqrt{\varepsilon^{-1}\log\left((K_{1}/\varepsilon)^{d}\right)}d\varepsilon\\
 & =\int_{0}^{1}\sqrt{d\varepsilon^{-1}(\log K_{1}-\log\varepsilon)}d\varepsilon\\
 & \leq\int_{0}^{1}\sqrt{(d\log K_{1})\varepsilon^{-1}}d\varepsilon+\sqrt{-d\varepsilon^{-1}\log\varepsilon}d\varepsilon<\infty.
\end{align*}

Therefore, it follows, by Theorem 8.3 of \cite{rio2017asymptotic},
that $\delta_{T}:=\sup_{(\gamma,u)\in\Gamma\times[c_{1},1-c_{2}]}|\hat{R}(\gamma,u)-R(\gamma,u)|=O_{P}(T^{-1/2})$. 

By the definition of $\gamma^{*}(u)$ and $\hat{\gamma}(u)$, we observe
that $R(\gamma^{*}(u),u)\leq R(\hat{\gamma}(u),u)\leq\hat{R}(\hat{\gamma}(u),u)+\delta_{T}\leq\hat{R}(\gamma^{*}(u),u)+\delta_{T}\leq R(\gamma^{*}(u),u)+2\delta_{T}$.
Hence, 
\begin{equation}
0\leq R(\hat{\gamma}(u),u)-R(\gamma^{*}(u),u)\leq2\delta_{T}.\label{eq: thm linear QR eq 4}
\end{equation}

For any $\gamma\in\Gamma$, we observe that 
\begin{align*}
 & E\left[\rho_{u}(Y_{t}-X_{t}^{\top}\gamma)-\rho_{u}(Y_{t}-X_{t}^{\top}\gamma^{*}(u))\mid X_{t}=x\right]\\
 & =\int f(y,x)\left(\rho_{u}(y-x^{\top}\gamma)-\rho_{u}(y-x^{\top}\gamma^{*}(u))\right)dy\\
 & \overset{\text{(i)}}{=}\int f(y,x)\left[-x^{\top}(\gamma-\gamma^{*}(u))\left(u-\oneb\{y-x^{\top}\gamma^{*}(u)\leq0\}\right)\right]dy\\
 & \qquad+\int f(y,x)\left[\int_{0}^{x^{\top}(\gamma-\gamma^{*}(u))}\left(\oneb\{y-x^{\top}\gamma^{*}(u)\leq s\}-\oneb\{y-x^{\top}\gamma^{*}(u)\leq0\}\right)ds\right]dy\\
 & =(F(x^{\top}\gamma^{*}(u),x)-u)x^{\top}(\gamma-\gamma^{*}(u))+\int_{0}^{x^{\top}(\gamma-\gamma^{*}(u))}\left(F(s+x^{\top}\gamma^{*}(u),x)-F(x^{\top}\gamma^{*}(u),x)\right)ds,
\end{align*}
where (i) follows by Equation (4.3) of \cite{koenker2005quantile}.
By the optimality condition of $\gamma^{*}(u)=\arg\min_{\gamma\in\Gamma}E\rho_{u}(Y_{t}-X_{t}^{\top}\gamma)$,
we have $E(F(X_{t}^{\top}\gamma^{*}(u),X_{t})-u)X_{t}^{\top}=0$.
Thus, the above display implies that for any $\gamma\in\Gamma$,
\begin{align*}
 & R(\gamma,u)-R(\gamma^{*}(u),u)\\
 & =E\left[\int_{0}^{X_{t}^{\top}(\gamma-\gamma^{*}(u))}\left(F(s+X_{t}^{\top}\gamma^{*}(u),x)-F(X_{t}^{\top}\gamma^{*}(u),x)\right)ds\right]\\
 & \overset{\text{(i)}}{\geq}\frac{1}{2}C_{2}E\left(X_{t}^{\top}(\gamma-\gamma^{*}(u))\right)^{2}\geq\frac{1}{2}C_{2}C_{3}\|\gamma-\gamma^{*}(u)\|_{2}^{2},
\end{align*}
where (i) follows by $f(y,x)\geq C_{2}$ for $y\in[s_{1}(x),s_{2}(x)]$.
By \eqref{eq: thm linear QR eq 4} and the above display, 
\[
\frac{1}{2}C_{2}C_{3}\|\hat{\gamma}(u)-\gamma^{*}(u)\|_{2}^{2}\leq R(\hat{\gamma}(u),u)-R(\gamma^{*}(u),u)\leq2\delta_{T}.
\]

Since this bound holds for any $u$, we have that 
\[
\sup_{u\in[c_{1},1-c_{1}]}\|\hat{\gamma}(u)-\gamma^{*}(u)\|_{2}^{2}\leq4\delta_{T}/(C_{2}C_{3}).
\]

Since we have proved $\delta_{T}=o_{P}(1)$, we have $\sup_{u\in[c_{T},1-c_{T}]}\|\hat{\gamma}(u)-\gamma^{*}(u)\|_{2}=o_{P}(1)$. 

\textbf{Step 2:} show the desired result.

Let $\varepsilon_{T}=\sup_{u\in[c_{T},1-c_{T}]}\|\hat{\gamma}(u)-\gamma^{*}(u)\|_{2}$.
Then 
\begin{equation}
\sup_{\|x\|_{2}\leq C_{1},u\in[c_{1},1-c_{1}]}|x^{\top}\hat{\gamma}(u)-x^{\top}\gamma^{*}(u)|\leq C_{1}\varepsilon_{T}.\label{eq: thm linear QR eq 9}
\end{equation}

We observe that for any $x\in\RR^{d}$ with $\|x\|_{2}\leq C_{1}$,
\[
\hF(y,x)=c_{T}+\int_{c_{T}}^{1-c_{T}}\oneb\{x^{\top}\hat{\gamma}(u)\leq y\}du\overset{\text{(i)}}{\leq}c_{T}+\int_{c_{T}}^{1-c_{T}}\oneb\{x^{\top}\gamma^{*}(u)-C_{1}\varepsilon_{T}\leq y\}du=F^{*}(y+C_{1}\varepsilon_{T},x),
\]
where (i) follows by \eqref{eq: thm linear QR eq 9}. Similarly, we
can show that $\hF(y,x)\geq F^{*}(y-C_{1}\varepsilon_{T},x)$. Therefore,
\[
|\hF(y,x)-F^{*}(y,x)|\leq\max\left\{ F^{*}(y+C_{1}\varepsilon_{T},x)-F^{*}(y,x),\ F^{*}(y,x)-F^{*}(y-C_{1}\varepsilon_{T},x)\right\} \leq h(C_{1}\varepsilon_{T}).
\]

Since this bounds holds for any $y$ and $x$, we have that $\sup_{\|x\|_{2}\leq C_{1}}\sup_{y\in\RR}|\hF(y,x)-F^{*}(y,x)|\leq h(C_{1}\varepsilon_{T})$.
By $\varepsilon_{T}=o_{P}(1)$ and $\lim_{\delta\rightarrow0}h(\delta)=0$,
the desired result follows. 
\end{proof}

\newpage

\section{Additional figures}
\label{app:additional_figures}

\begin{figure}[H]
\begin{center}
\includegraphics[width=0.325\textwidth,trim=0 0cm 0 0.5cm]{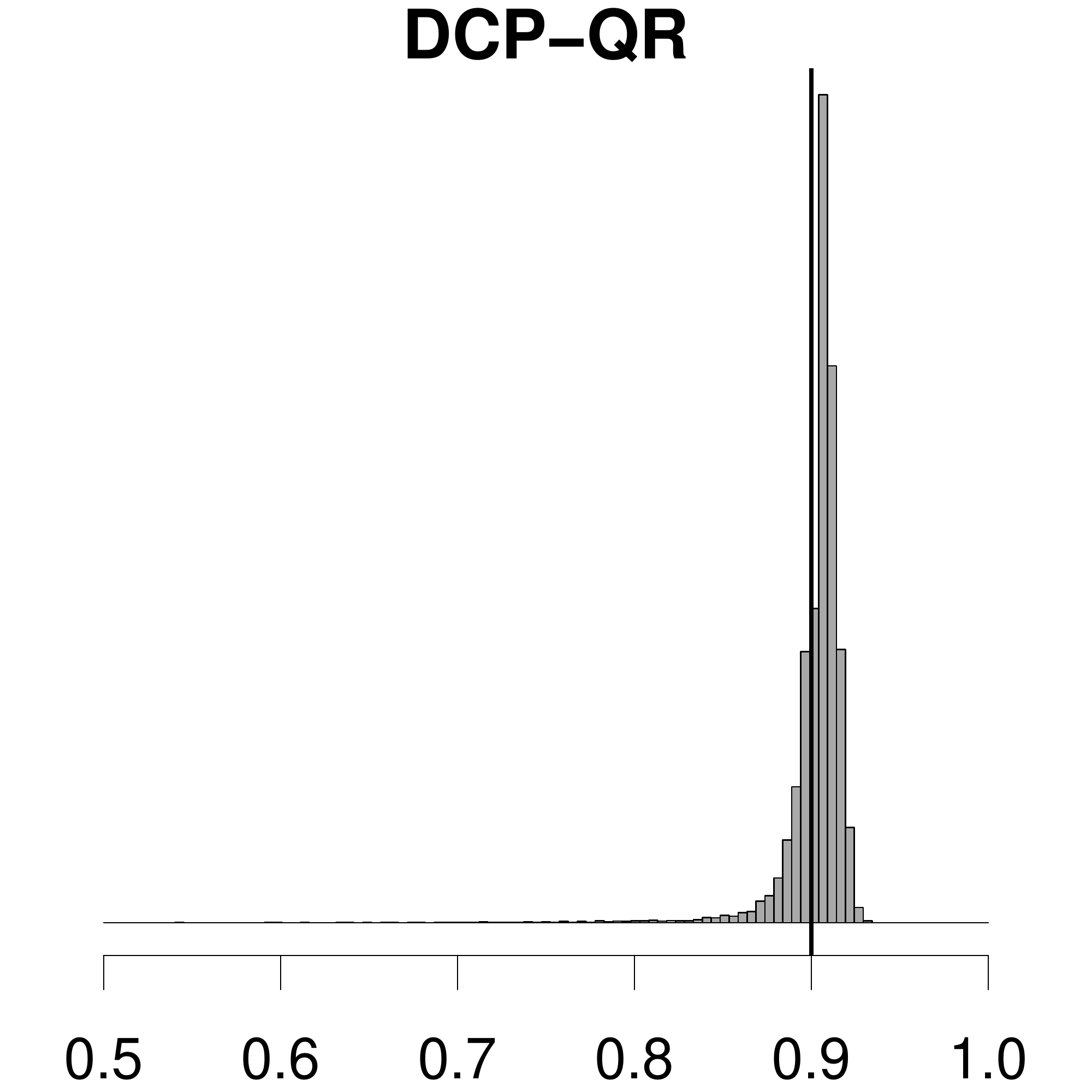}
\includegraphics[width=0.325\textwidth,trim=0 0cm 0 0.5cm]{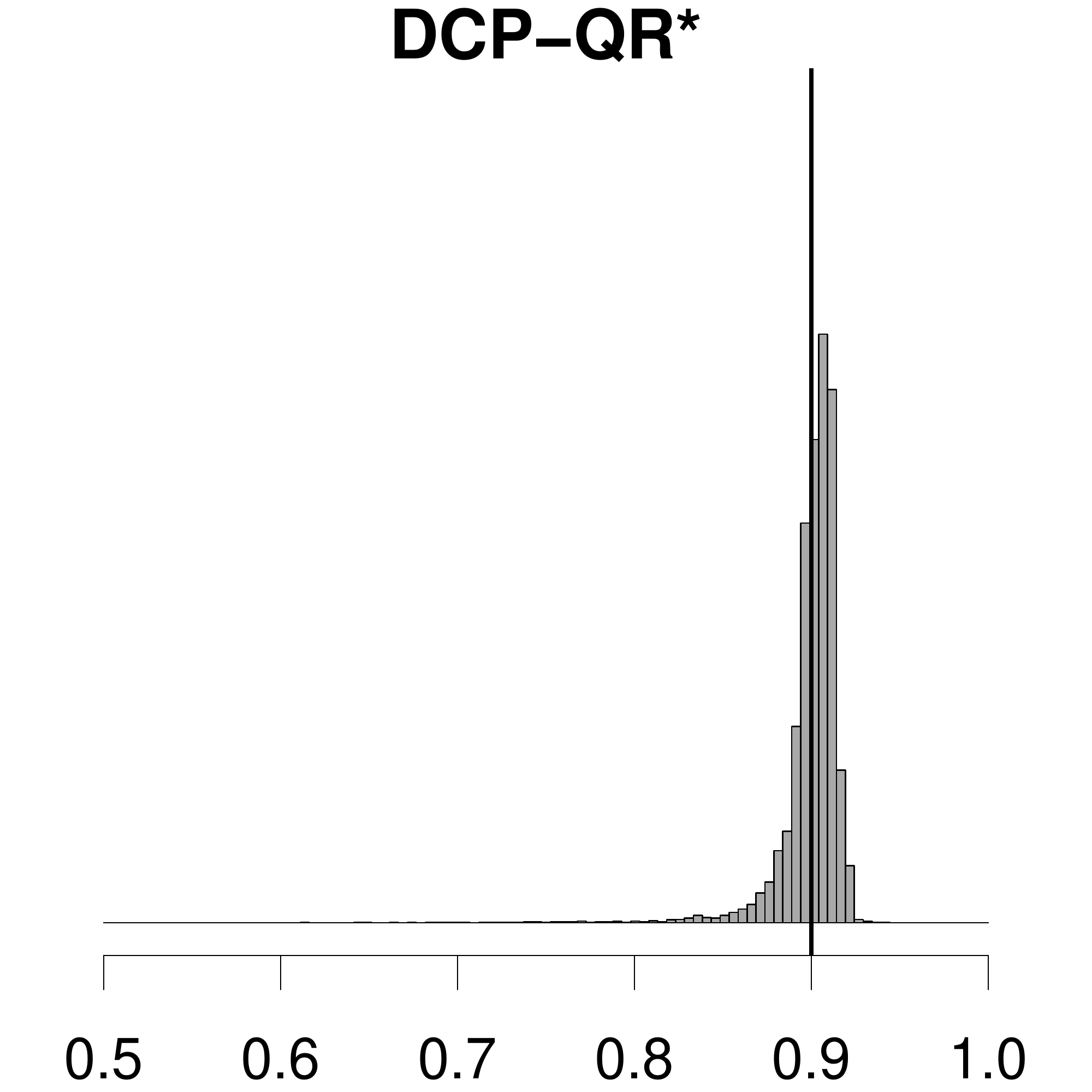}
\includegraphics[width=0.325\textwidth,trim=0 0cm 0 0.5cm]{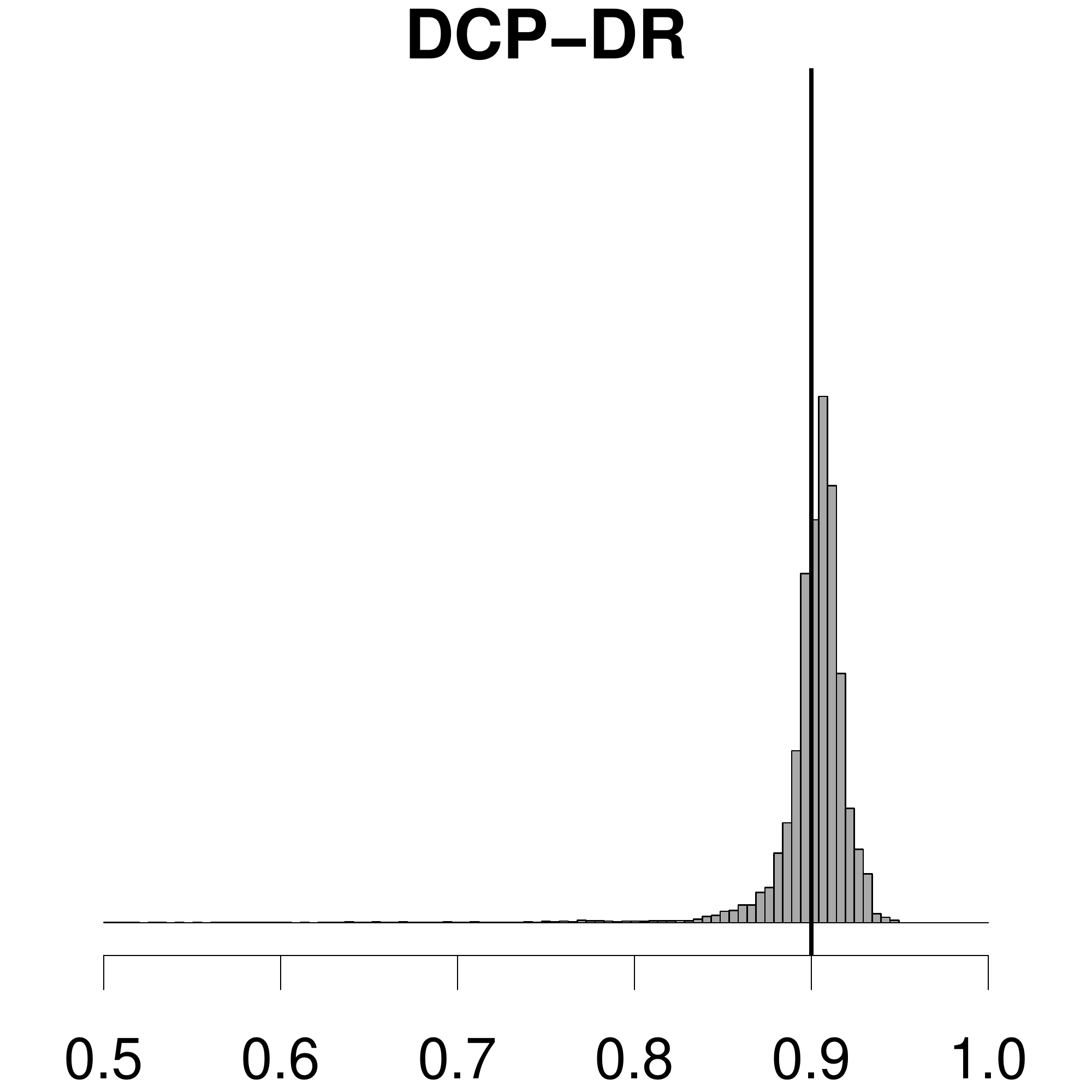}

\includegraphics[width=0.325\textwidth,trim=0 0cm 0 0]{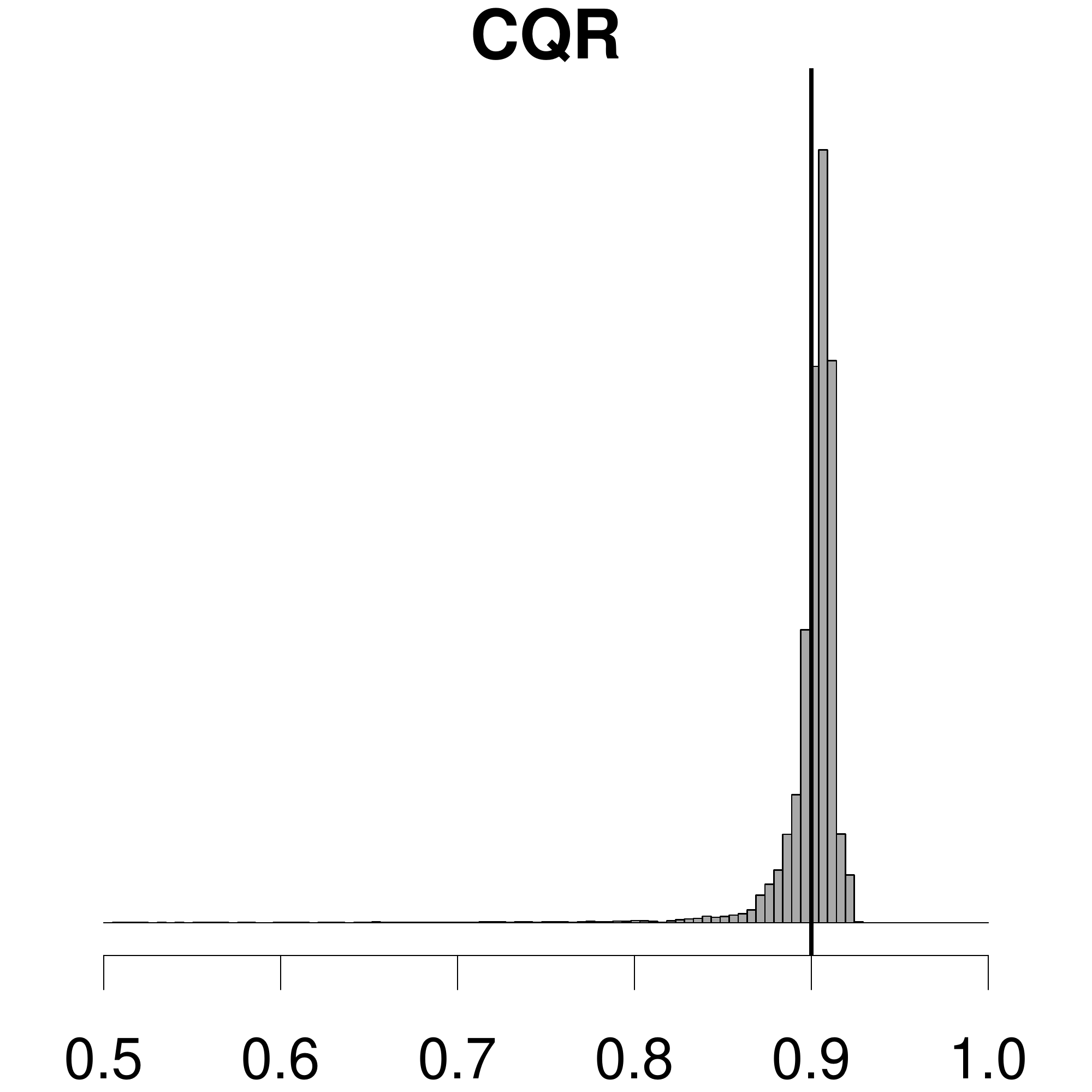}
\includegraphics[width=0.325\textwidth,trim=0 0cm 0 0]{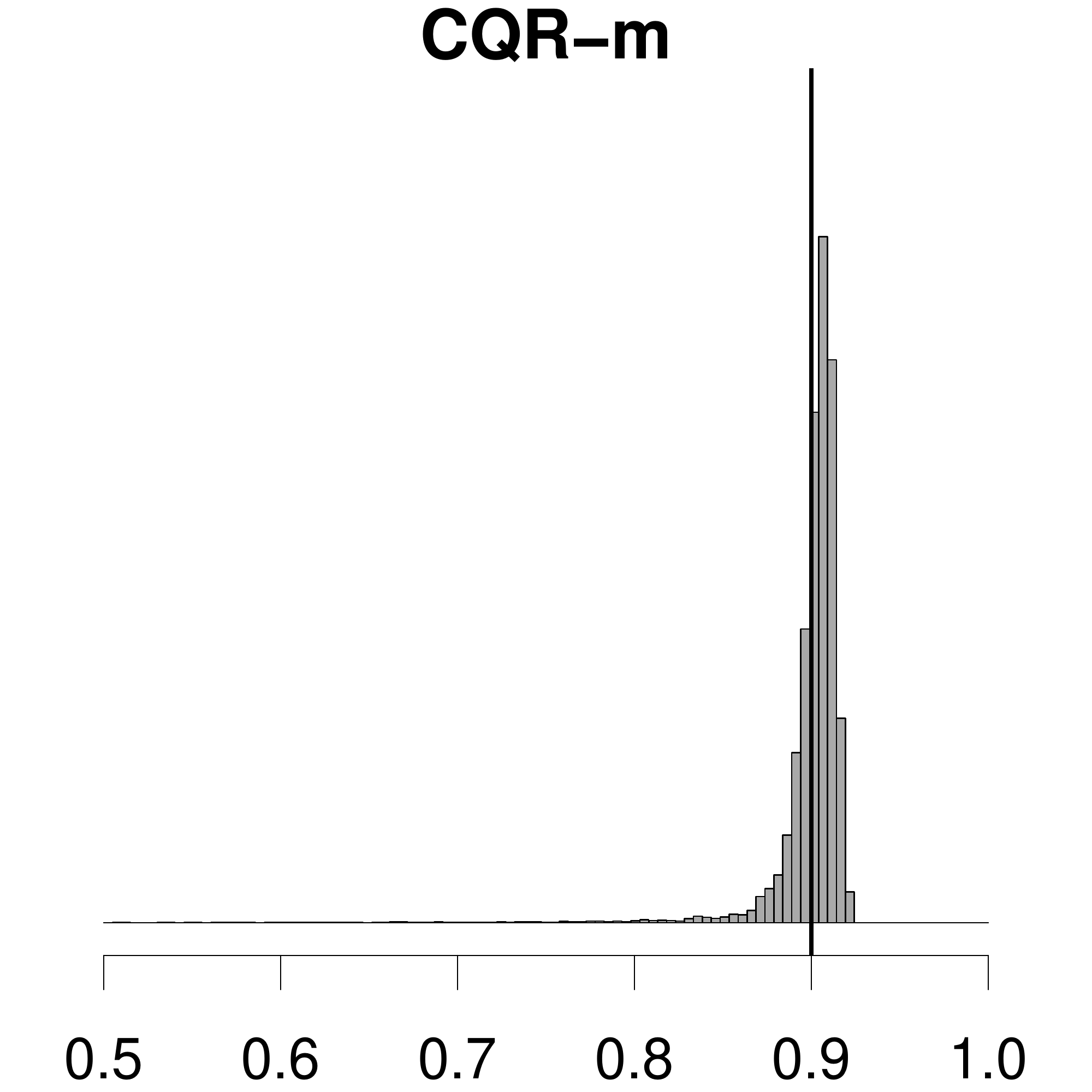}
\includegraphics[width=0.325\textwidth,trim=0 0cm 0 0]{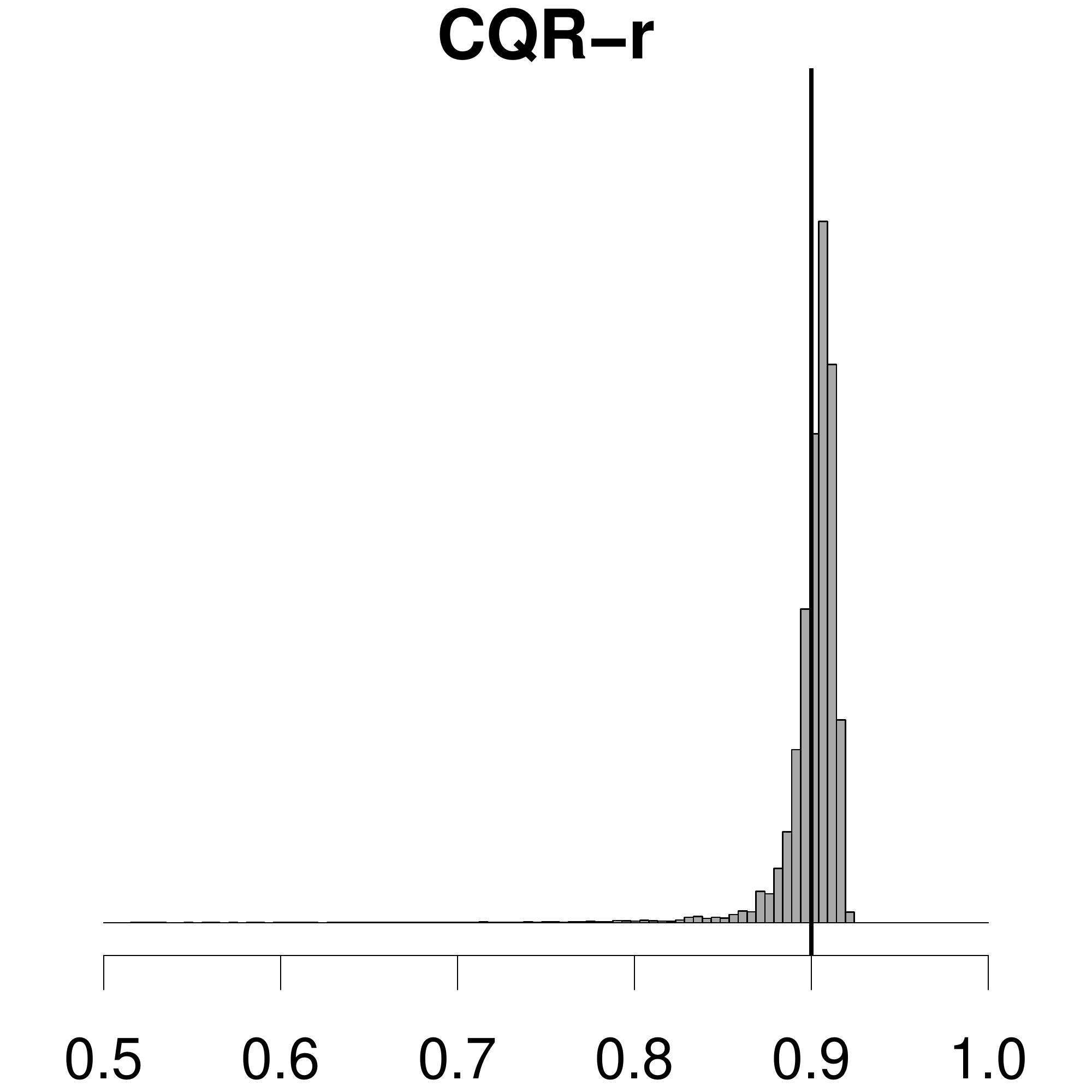}

\includegraphics[width=0.325\textwidth,trim=0 0cm 0 0]{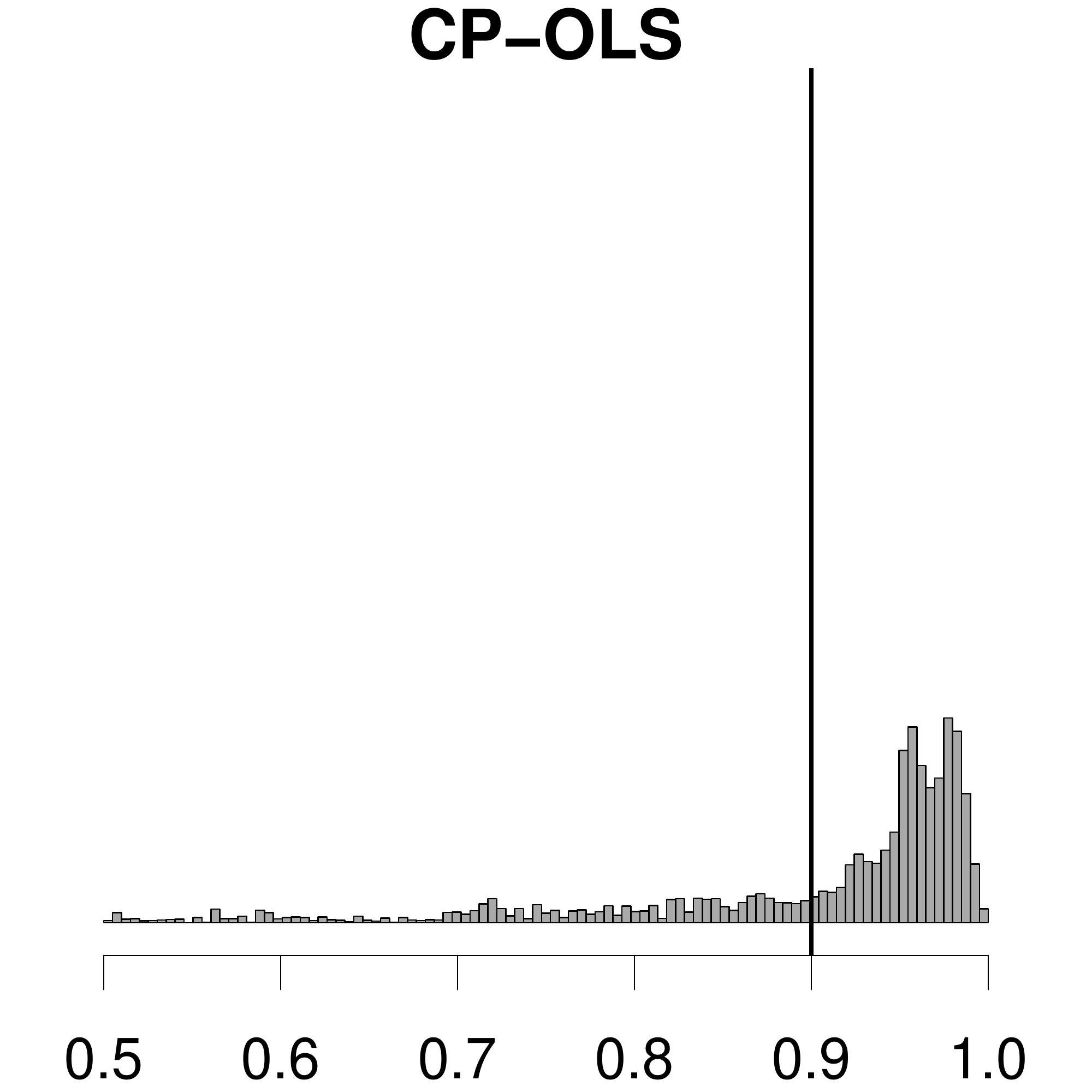}
\includegraphics[width=0.325\textwidth,trim=0 0cm 0 0]{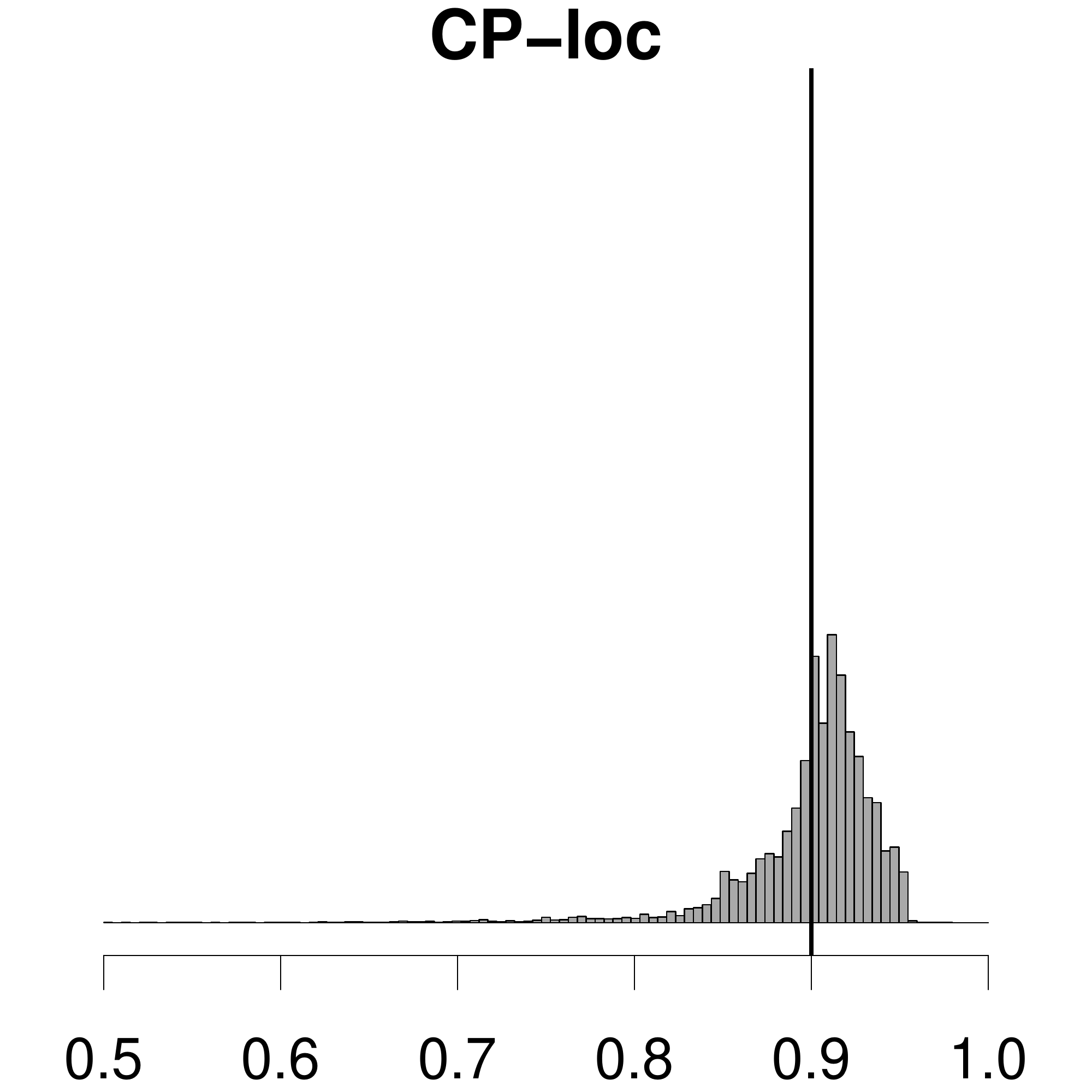}

\caption{Histograms of estimated conditional coverage probability. Vertical line at nominal coverage of $1-\alpha=0.9$.}
\label{fig:conditional_coverage_cps}
\end{center}
\end{figure}

\end{document}